\documentclass[english,12pt,aps,prd,a4paper,preprintnumbers,floatfix,nofootinbib,superscriptaddress, notitlepage]{revtex4} 
\usepackage[usenames,dvipsnames]{color}  
\usepackage{graphicx}
\usepackage{setspace}
\usepackage{caption}
\usepackage{subcaption}
\usepackage{amsmath}
\usepackage{amssymb}
\usepackage[colorlinks=true,citecolor=darkred,urlcolor=darkred, pdfborder={0 0 0}]{hyperref}
\usepackage[normalem]{ulem}
\usepackage{xcolor}
\usepackage{tikz-feynman,contour}
\usetikzlibrary{arrows}
\tikzstyle{connector} = [draw, -latex']
\usepackage{float}

\makeatletter
\def\p@subsection{}
\makeatother

\definecolor{darkred}{rgb}{0.6,0,0}

\definecolor{linkcolor}{rgb}{0,0,0.5}

\def\gsim{\raise0.3ex\hbox{$\;>$\kern-0.75em\raise-1.1ex\hbox{$\sim\;$}}}
\def\lsim{\raise0.3ex\hbox{$\;<$\kern-0.75em\raise-1.1ex\hbox{$\sim\;$}}}

\def\beqn#1{\begin{equation}\label{#1}}
\def\eeqn{\end{equation}}

\def\beqa#1{\begin{eqnarray}\label{#1}}
\def\eeqa{\end{eqnarray}}

\def\Z2{$\mathcal{Z_2}$}

\newcommand {\ignore}[1]{}

\def\321{$\mathrm{SU(3) \otimes SU(2) \otimes U(1)}$ }

\newcommand{\AddrCFTP}{Departamento de F\'{\i}sica and CFTP, Instituto Superior T\'ecnico, Universidade de Lisboa, Av. Rovisco Pais 1, 1049-001 Lisboa, Portugal}
 
 \newcommand{\AddrIISERB}{Department of Physics,
 Indian Institute of Science Education and Research - Bhopal \\
 Bhopal Bypass Road, Bhauri, Bhopal, India}

\begin{document}

\title{\color{BrickRed} $h \to \Upsilon \gamma$ Decay: Smoking Gun Signature of Wrong-Sign $hb\bar{b}$ Coupling }
\author{Aditya Batra}\email{aditya.batra@tecnico.ulisboa.pt}
\affiliation{\AddrIISERB}
\affiliation{\AddrCFTP}
\author{Sanjoy Mandal}\email{smandal@kias.re.kr}
\affiliation{Korea Institute for Advanced Study, Seoul 02455, Korea}
\author{Rahul Srivastava}\email{rahul@iiserb.ac.in}
\affiliation{\AddrIISERB}

\begin{abstract}
  \vspace{1cm} 
  
We perform a model-independent study of new physics effects in the Higgs decay
$h \to \Upsilon \gamma$, focusing on scenarios that spoil the accidental cancellation
between the direct and indirect amplitudes. After imposing all existing constraints
from Higgs production and decay measurements, we find that a wrong-sign $h b\bar b$
coupling is the only viable scenario capable of enhancing the
$h \to \Upsilon \gamma$ decay width by nearly two orders of magnitude. Therefore,
an observation of a significantly enhanced $h \to \Upsilon \gamma$ rate at the LHC
or future colliders would provide unambiguous evidence for a wrong-sign
$h b\bar b$ coupling, directly pointing to the presence of an extended Higgs sector.

\end{abstract}
\maketitle


\section{INTRODUCTION}
\label{ref:sec1}
The Standard Model (SM) of particle physics is a theory that describes electromagnetic, weak and strong interactions using symmetry groups $SU(3)_C \otimes SU(2)_L \otimes  U(1)_Y $. It also consists of a framework that classifies all known fundamental particles with specific charges under these groups.  
Although it is an extremely successful theory with most of its predictions  confirmed by experiments, it still remains a flawed theory with  severe shortcomings. 
For example, it does not contain any particles that could be viable candidates for dark matter, as expected by observational cosmology~\cite{Planck:2018vyg}.	It also does not incorporate masses for neutrinos deduced from neutrino oscillation experiments~\cite{Kajita:2016cak,McDonald:2016ixn,KamLAND:2002uet,K2K:2002icj}. In addition, the SM also has several other shortcomings such as Higgs vacuum stability, hierarchy problem and baryon-antibaryon asymmetry, to name a few.   In order to overcome these limitations, one has to extend the particle content of the SM. The new particles and interactions present in such Beyond Standard Model (BSM) proposals can then be tested directly or indirectly in various currently running experiments such as at the Large Hadron Collider (LHC).

On the experimental front, in 2012, the LHC discovered a new spin-0 particle \cite{ATLAS:2012yve,CMS:2012qbp}. So far the experimental data (see Table \ref{tab:mu}) on decay channels and branching ratios (BR) is consistent with the predictions of the SM Higgs ($h$) boson. However, one must now do precision measurements of its couplings to other SM particles and verify whether there are any deviations from the predictions for the SM Higgs boson. Therefore, measuring the yet unmeasured couplings and improving the precision of the measured ones is one of the main physics goals of not only Run3 of the LHC but also of various proposed future colliders \cite{Behnke:2013xla, CidVidal:2018eel, FCC:2018byv,FCC:2018evy,FCC:2018vvp,FCC:2018bvk}.  

Among the various Higgs couplings, measuring the Yukawa coupling of Higgs to fermions of the SM is especially challenging. This is because in the SM as well as in many simple BSM models, the Yukawa coupling strength is directly proportional to the mass of the fermion the Higgs couples with.  Since the light quarks ($u$, $d$, $s$) and charged leptons ($e$, $\mu$)  have very small masses,  measuring their Yukawa couplings through Higgs decay to them becomes very challenging as the branching ratios are very small (see Figure \ref{fig:h-to-xx} and Table \ref{tab:mu}). The branching ratios of Higgs decay to $b$ and $c$ quarks as well as to the charged lepton $\tau$ are not small but measuring them is still challenging due to complications arising from background events as well as difficulties in correctly tagging the resulting jets. As a result, only a few such decay modes have been measured at the LHC, as shown in Table \ref{tab:mu}. Thus, before one can conclude that the $125$ GeV scalar particle observed at the LHC is the SM Higgs boson, one needs to measure and compare these Higgs fermionic decay modes to the SM predictions.

Among all the Yukawas, perhaps  the Higgs coupling to b-quark is of particular interest. In the SM, the $h \to b \bar{b}$ decay has the largest branching ratio and has also been measured at the LHC \cite{ATLAS:2018kot,CMS:2018nsn}. Despite having the largest branching ratio, the observation of the $h \to b \bar{b}$ decay mode was very challenging and took quite some time. The current measured value is consistent with the SM prediction but has large error bars, leaving considerable room for deviations from the SM value due to presence of new physics. Even if the uncertainty in measurement is reduced substantially in the future, the $h b \bar{b}$ coupling can still have BSM imprints on it. One such simple but interesting scenario is the possibility of the $h b \bar{b}$ coupling being of the ``wrong-sign''.  

The wrong-sign  $h b \bar{b}$ coupling can in fact arise in some of the natural and popular BSM extensions with extended scalar sectors. For example, the Two-Higgs Doublet Model (2HDM) \cite{Deshpande:1977rw,Branco:2011iw} is one of the simplest conceivable scalar extensions of the Standard Model. The 2HDMs arise in various contexts, with the scalar sector of many models being of the 2HDM form. For instance, the minimal versions of SuperSymmetry such as the MSSM have a specific (type-II) 2HDM scalar sector and such is also the case in the Scotogenic model for neutrino mass and dark matter \cite{Ma:2006km}. Furthermore, 2HDMs can generate baryon asymmetry of substantial size in the Universe due to new sources of  CP-violation through the complex parameters in the potential. There also exist 2HDMs that contain viable dark matter candidates such as Inert 2HDMs. Needless to say, 2HDMs are not only popular but also ubiquitous among BSM models. Even among the simplest 2HDM scenarios  with an additional $\mathbb{Z}_2$ symmetry,  the wrong-sign solution exists in the Type II and Flipped 2HDMs where the sign of the $ h b \bar{b} $ coupling can be opposite to that of the SM. 

Given the possibility of the wrong-sign solution among BSM physics, how can we look for its phenomenological signatures? The effect of the wrong sign must be examined indirectly because a sign change has no influence on the $h \to b \bar{b}$ rate. This means that the ideal place to look for wrong-sign solutions is in processes where two or more Feynman diagrams contribute. In such processes the interference term will preserve some information about any anomalous and/or  wrong-sign $ h b \bar{b} $ coupling \cite{Ferreira:2014naa,Fontes:2014tga,Fontes:2014xva,Modak:2014ywa}. The aim of this work is to conclusively show that the rare Higgs decay $h \to \Upsilon \gamma$ is the ideal place to look for it\footnote{Throughout this work we will mostly concentrate on the decay mode $h \to \Upsilon (1S) \, \gamma$, however the results and the conclusion of this work can be easily generalized and applied to $h \to \Upsilon (2S)\, \gamma$ and $h \to \Upsilon (3S)\,  \gamma$ decay modes as well. For the sake of brevity, we have also dropped the $(1S)$ label and will simply denote  it as $h \to \Upsilon \gamma$ decay.} and observation of such a decay by the LHC or any future collider with a branching ratio substantially larger than the SM expectation will be a smoking gun signature of wrong-sign $ h b \bar{b} $ coupling.

In the SM, the decay $h \to \Upsilon \gamma$ comprises of two types  of Feynman diagrams called the direct and indirect diagrams, see Fig. \ref{fig:h-to-upsilon}. In the SM, the direct and indirect diagrams are of  about equal magnitude but interfere destructively. Due to the cancellation between the two diagrams \cite{Bodwin:2013gca,Konig:2015qat,Zhou:2016sot,Brambilla:2019fmu}, the decay width of $h \to \Upsilon \gamma$ in the SM is greatly reduced when compared, for example, to $h \to J/\psi\, \gamma$, see Table \ref{tab:rare} in Appendix~\ref{sec:measure}. This accidental cancellation in 
$h \to \Upsilon \gamma$ in the SM makes it an ideal channel to look for BSM physics. Any new physics contribution which can disrupt the accidental cancellation can potentially enhance the $h \to \Upsilon \gamma$ branching ratio significantly. Indeed it was shown in Ref. \cite{Modak:2016cdm, Modak:2018zro} that the wrong-sign  
$ h b \bar{b} $ coupling can enhance the $h \to \Upsilon \gamma$ branching ratio by about two orders of magnitude. 
Therefore, $h \to \Upsilon \gamma$ is the cleanest decay mode for probing the wrong-sign solution \cite{Modak:2016cdm}. 

However, the reverse question still remains, i.e., if one observes the $h \to \Upsilon \gamma$ branching ratio to be significantly enhanced compared to the SM prediction, can one conclusively say that it is indeed due to the wrong-sign 
$ h b \bar{b} $ coupling? The answer is not so straightforward as there is also the possibility of the cancellation being broken due to changes in the sign or magnitude of the indirect contribution. Therefore, one must thoroughly and systematically analyze all possible type of new physics contributions that can disrupt this accidental cancellation  by changing the contribution coming from indirect diagram(s). In this work, we perform such a model independent analysis taking into account all possible contributions that can change either the sign or the magnitude of the direct or indirect contribution and thus break the accidental cancellation of the SM. After considering all such possibilities, we show that the wrong-sign $h b \bar{b} $ coupling is the only way one can change the branching ratio of the  $h \to \Upsilon \gamma$ decay significantly. We can thus conclusively say that if the LHC or any future collider observes $h \to \Upsilon \gamma$ to be much larger than the SM expectation then this is not only a conclusive signature of new physics but also that this new physics has to be related with the wrong-sign $h b \bar{b} $ coupling.

This paper is organized as follows. In section \ref{sec:htoUpG}, we introduce the rare Higgs decay to $ \Upsilon\gamma $. In sections \ref{sec:direct} and \ref{sec:indirect}, we discuss the modifications to the direct and indirect contributions of the $ h\to\Upsilon\gamma $ decay respectively, looking in a model independent way, at all possible new particle contributions to the indirect mode in detail. Finally, we summarize our results in section \ref{sec:conclusions}. 
Further details on the Higgs boson couplings and decay channels, including their current experimental status, as well as the formulas relevant to our calculations, are provided in the appendices.

\section{Higgs decay to $\Upsilon \gamma$}
\label{sec:htoUpG}

\begin{figure}[!htbp]
	\centering
	\begin{subfigure}{0.49\linewidth}
		\centering
		\includegraphics[width=0.6\linewidth]{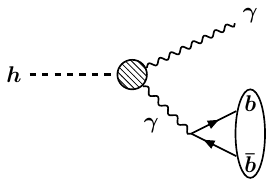}
		\caption{\centering}
		\label{indirect}
	\end{subfigure}
	\begin{subfigure}{0.49\linewidth}
		\centering
		\includegraphics[width=0.6\linewidth]{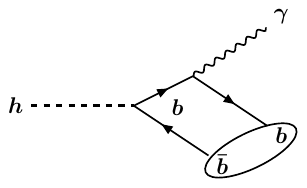}
		\caption{\centering}
		\label{direct}
	\end{subfigure}
		\caption{Indirect (left) and direct (right) Feynman diagrams of the $ h\rightarrow\Upsilon\gamma $ process.}
		\label{fig:h-to-upsilon}
\end{figure}

The $ h \to \Upsilon (1S, 2S, 3S) \, \gamma $ decay process has two leading contributions from the so called  direct and indirect Feynman diagrams shown in Fig. \ref{fig:h-to-upsilon}. 
In the indirect mode, $ h $ decays to $\gamma\gamma$ via a loop and one photon  then morphs into an $\Upsilon$, as shown in Fig.~\ref{indirect}. In the SM, the primary contribution to the $ h\to\gamma\gamma $ loop comes from the $ W $ boson followed by the top-quark which interferes destructively. In the direct mode, $ h $ decays to $ b \bar{b}$ and either $ b $ or $ \bar{b} $ emits a photon, as shown in Fig.~\ref{direct}. Note that in SM there is another indirect diagram in which the virtual $\gamma$ is replaced by a virtual $ Z $ boson  which morphs into an $\Upsilon$.  However its contribution is negligible due to the large mass of the $Z$ boson.

The $h \to \Upsilon (1S) \gamma$ decay width in the SM is given as~\cite{Bodwin:2013gca,Konig:2015qat,Zhou:2016sot,Brambilla:2019fmu}:
\begin{equation}
\Gamma[h \to \Upsilon \gamma]
=\frac{1}{8 \pi} \frac{m_h^2 - m_\Upsilon^2}{m_h^2}
\left| {\cal A}_\textrm{direct} + {\cal A}_\textrm{indirect} \right|^2 ,
\end{equation}
where the direct and indirect decay amplitudes are as given in \cite{Bodwin:2013gca}:
\begin{align}
{\cal A}_\textrm{direct}
&=
-  K\, \frac{2}{\sqrt{3}} e\, 
\left( \sqrt{2} G_F \frac{m_\Upsilon}{m_h}\right)^{1/2}
\frac{m_h^2- m_\Upsilon^2}{(m_h^2- m_\Upsilon^2/2 - 2 m_b^2)}\
\phi_0(\Upsilon),
\nonumber\\*[2mm\,]
{\cal A}_\textrm{indirect}
&=
-\frac{e\, g_{\Upsilon \gamma}}{m_\Upsilon^2}
\left( \sqrt{2} G_F \right)^{1/2} \frac{\alpha}{\pi}
\frac{m_h^2- m_\Upsilon^2}{\sqrt{m_h}} \zeta,
\label{amps}
\end{align}
where $G_F$ is the Fermi constant, $e$ is the absolute charge of the electron, $m_\Upsilon$ and $m_b$ are the $\Upsilon$ and $b$-quark masses, $\alpha$ is the fine-structure constant, $\phi_0^2(\Upsilon) \sim 0.512$ $\textrm{GeV}^3$ is the wave function of $\Upsilon$ at the origin, $g_{\Upsilon \gamma} = \frac{2}{\sqrt{3}} \sqrt{m_\Upsilon}\, \phi_0(\Upsilon)$, $\zeta$ is the effective $h \gamma \gamma$ coupling at one-loop and $ K = 0.689 $ is an added factor that comes from NLO corrections~\cite{Bodwin:2013gca}. Note that the sign of $\zeta$ is negative which implies that the overall sign of the indirect amplitude is positive and that of the direct amplitude is negative.
This means that the direct and indirect contributions in the SM interfere destructively. Moreover, in the SM it so happens that the contributions of the direct and indirect diagrams is almost of the same magnitude. Thus, there is an almost complete cancellation of the two contributions leading to a much smaller decay width compared to other Higgs quarkonium decays listed in Table \ref{tab:rare}. Before moving on, note that due to the close nature of the cancellation, the SM branching ratio of $h \to \Upsilon \gamma$ is extremely sensitive to the values of the input parameters and different groups have reported slightly different values of the SM branching ratios\cite{Bodwin:2013gca,Konig:2015qat,Zhou:2016sot,Brambilla:2019fmu}. However, since in our work we are interested in new physics effects which can change the branching ratio by orders of magnitude, the high sensitivity to input parameters will not have any appreciable effect on our results and conclusions.  

The accidental cancellation between the direct and indirect diagrams for $h \to \Upsilon \gamma$ in the SM makes this decay channel an excellent hunting ground for BSM models. Any new physics which disrupts this accidental cancellation will increase the branching ratio of $ h \to \Upsilon\gamma $, potentially by orders of magnitude.
This disruption can happen primarily because of either (a) Modifications in the Direct Diagram or (b) Modification in the Indirect Diagram. In the next sections, we will look at these possible modifications.
\section{Possible modifications to the direct diagram}
\label{sec:direct}

We begin with first looking at the possible modifications to the direct diagram. The only way the direct diagram of $ h \to \Upsilon\gamma $ can be modified is if the $ h b \bar{b} $ vertex is modified. The other vertex $ b \bar{b} \gamma $ being a QED vertex is highly constrained, with no scope for any significant modification.
The $h b \bar{b}$ coupling can be modified from its SM value in many new physics scenarios. Such  modifications can lead to an ``anomalous'' $h b \bar{b}$ coupling. More interestingly, in certain extensions of the SM with additional scalar fields, the sign of the $ h b \bar{b} $ coupling is opposite to that of the SM. This is called the ``wrong-sign'' solution.
As shown in \cite{Modak:2016cdm},  this effect produces a very significant modification to the direct diagram of $ h \to \Upsilon\gamma$. The simplest models in which this occurs are the Type II and Flipped versions of the  Two Higgs Doublet Models. In the following subsections, we will briefly  look at the 2HDM and its wrong-sign solution. The final analysis we present for the impact of the wrong-sign solution on the direct diagram of $ h \to \Upsilon\gamma $ will be general and is applicable to  any theory with a wrong-sign $ h b \bar{b} $ coupling.

\subsection{The Two Higgs Doublet Model}

In the 2HDM, the most general potential can be written as \cite{Gunion:2002zf,Branco:2011iw}:
\begin{align}
\label{V2HDM}
V = &-\mu^2_{1} (\Phi_1^\dagger \Phi_1) -\mu^2_2 (\Phi_2^\dagger \Phi_2)
-\biggl[ \mu^2_{12}(\Phi_1^\dagger \Phi_2) + {\rm{h.c.}} \biggr]
\\
&
+ \lambda_{1} (\Phi_1^\dagger \Phi_1)^2
+ \lambda_{2} (\Phi_2^\dagger \Phi_2)^2
+ \lambda_{3} (\Phi_1^\dagger \Phi_1)(\Phi_2^\dagger \Phi_2)
+ \lambda_{4} (\Phi_1^\dagger \Phi_2)(\Phi_2^\dagger \Phi_1)
\nonumber\\
&+ \biggl[\lambda_{5} (\Phi_1^\dagger \Phi_2)^2
+\lambda_{6}  (\Phi_1^\dagger \Phi_1) (\Phi_1^\dagger \Phi_2)
+\lambda_{7}  (\Phi_2^\dagger \Phi_2) (\Phi_1^\dagger \Phi_2)
+ {\rm{h.c.}} \biggr].\nonumber
\end{align} 
The Higgs doublet fields can be parametrized as \cite{Aoki:2009ha}:
\begin{align*}
\Phi_i=\begin{pmatrix}\phi_i^+\\\frac1{\sqrt2}(v_i+h_i-i\,\eta_i)
\end{pmatrix},\hspace{1cm} i=1,2
\end{align*}
where $ v_i $ is the vacuum expectation value (vev) of $ \Phi_i $, $\sqrt{v_1^2+v_2^2}=v=(\sqrt{2} G_F)^{-1/2}\simeq 246$ GeV, $ h_i $, $ \eta_i $ are real scalars, and $ \phi_i^+ $ is a complex scalar.
The mass eigenstates are defined by:
\begin{align*}
\begin{pmatrix}h_1\\h_2\end{pmatrix}=\text{R}(\alpha)
\begin{pmatrix}H\\h\end{pmatrix},\quad
\begin{pmatrix}\eta_1\\\eta_2\end{pmatrix}=\text{R}(\beta)
\begin{pmatrix}G^0\\A\end{pmatrix},\quad
\begin{pmatrix}\phi_1^+\\\phi_2^+\end{pmatrix}=\text{R}(\beta)
\begin{pmatrix}G^+\\H^+\end{pmatrix},
\end{align*}
where $ \text{R}(\theta) $ is the rotation matrix:
\begin{align}
\text{R}(\theta)=\begin{pmatrix}\cos\theta&-\sin\theta\\
\sin\theta&\cos\theta\end{pmatrix}.
\label{Rotation}
\end{align}
There are five physical scalar particles: two CP-even scalars $h$ and $H$, one CP-odd scalar $A$, and a pair of charged scalars $H^\pm$. $G^0$ and $G^\pm$ are the Goldstone bosons. $\alpha$ and $\beta$ are mixing angles ($\tan\beta=v_2/v_1$). Here we take the lightest scalar ($h$) to be the 125 GeV scalar particle found at the LHC and will simply refer to it as the Higgs boson.
The coupling of $h$ with gauge bosons are:
\begin{align}
{\cal L}_{hVV} &=
\sin{(\beta - \alpha)} h
\left[
\frac{m_Z^2}{v} Z^\mu Z_\mu
+ 2\, \frac{m_W^2}{v} W^{+ \mu} W^-_\mu
\right].
\end{align}
If one takes the limit $\sin{(\beta - \alpha)}=1$ then the coupling of $h$ with the gauge bosons will be same as the SM Higgs coupling. This limit is often called the ``alignment'' limit in literature.

Since fermions can couple to any of the two scalar doublets, 2HDMs allow tree-level Flavour Changing Neutral Currents (FCNCs) which are restricted experimentally. These interactions can be forbidden by introducing discrete symmetries like $\mathbb{Z}_2$. If both doublets have different $\mathbb{Z}_2$ charges, then fermions couple to only one doublet depending on their own $\mathbb{Z}_2$ charges.
Imposing the $\mathbb{Z}_2$ symmetry forbids the $ \lambda_{6}  (\Phi_1^\dagger \Phi_1) (\Phi_1^\dagger \Phi_2) $ and $ \lambda_{7}  (\Phi_2^\dagger \Phi_2) (\Phi_1^\dagger \Phi_2) $ terms in the potential. The $\mu_{12}$ which also breaks the $\mathbb{Z}_2$ is usually allowed in the potential, thus breaking the $\mathbb{Z}_2$ symmetry softly.

Depending on the $\mathbb{Z}_2$ charges of the fermions, there are four types of 2HDMs without tree-level FCNCs: Type-I, Type-II, Lepton-specific and Flipped as shown in Table \ref{tab:2hdms}.
\begin{table}[!htbp]
	\begin{center}
		\begin{tabular}{|c|c|c|c|c|}  \hline
			& \hspace{0.1cm}Type I  \hspace{0.1cm}   &\hspace{0.1cm} Type II \hspace{0.1cm}& \hspace{0.1cm}Lepton-specific \hspace{0.1cm}& \hspace{0.1cm}Flipped  \hspace{0.1cm}   \\
			\hline
			\hspace{0.3cm}$ u $  \hspace{0.3cm}   & $\Phi_2$   &$ \Phi_2$ & $\Phi_2$  &$ \Phi_2$    \\
			\hline
			$ d $     & $\Phi_2$       &$\Phi_1$    &$ \Phi_2$       & $\Phi_1$    \\
			\hline
			$ l $     &$\Phi_2$    & $\Phi_1 $         & $\Phi_1$  &$ \Phi_2$  \\
			\hline
		\end{tabular}
	\end{center}
	\caption{The scalar doublets coupling to up-type quarks ($ u $), down-type quarks ($ d $) and leptons ($ l $) in the 2HDM models.}
	\label{tab:2hdms}
\end{table}
In the type-I 2HDM, all quarks and leptons couple to $\Phi_2$. In the type-II 2HDM, up-type quarks couple to $\Phi_2$, while down-type quarks and leptons couple to $\Phi_1$. In the lepton-specific 2HDM all quarks couple to $\Phi_2$,	while leptons couple to $\Phi_1$ while in the flipped 2HDM, up-type quarks and leptons couple to $\Phi_2$, while down-type quarks couple to $\Phi_1$.
 
The couplings of $ h $ to fermions are given by:
\begin{align}
\mathcal{L}_\text{yukawa} =
&-\sum_{f=u,d,\ell} \frac{m_f}{v}\xi_f\overline{f}fh,
\end{align}
where the factors $\xi_{u,d,l}$ are given in Table \ref{2HDM Couplings}.

\begin{table}[!htbp]
	\begin{center}
		\begin{tabular}{|c|c|c|c|c|}  \hline
{} &\hspace{0.1cm} Type I\hspace{0.1cm}  &\hspace{0.1cm} Type II \hspace{0.1cm}& \hspace{0.1cm}Lepton-specific\hspace{0.1cm} &\hspace{0.1cm} Flipped \hspace{0.1cm}    \\
\hline
\hspace{0.3cm}$\xi_u $\hspace{0.3cm}  &\hspace{0.1cm} $\cos\alpha/\sin\beta$ \hspace{0.1cm}  & \hspace{0.1cm}$\cos\alpha/\sin\beta$\hspace{0.1cm} &\hspace{0.1cm} $\cos\alpha/\sin\beta$ \hspace{0.1cm}& \hspace{0.1cm}$\cos\alpha/\sin\beta$  \hspace{0.1cm}    \\
\hline
$\xi_d $    & $\cos\alpha/\sin\beta$        &\hspace{0.1cm} $-\sin\alpha/\cos\beta$   \hspace{0.1cm}         & $\cos\alpha/\sin\beta$  & \hspace{0.1cm}$-\sin\alpha/\cos\beta$ \hspace{0.1cm}  \\
\hline
$\xi_\ell $ & $\cos\alpha/\sin\beta$     & $-\sin\alpha/\cos\beta$          & $-\sin\alpha/\cos\beta$   & $\cos\alpha/\sin\beta$   \\
\hline
\end{tabular}
\end{center}
\caption{Yukawa couplings of various fermions with the physical Higgs boson $ h $ in the four different types of 2HDM \cite{Branco:2011iw}.}
	\label{2HDM Couplings}
\end{table}

\subsection{Wrong-sign $h b \bar{b}$ coupling}

As mentioned before, the wrong-sign solution occurs in the Type II and Flipped 2HDMs. Let us look at its origin in a bit of detail. As shown in Table \ref{2HDM Couplings}, in Type II and Flipped 2HDM, the coupling of $ h $ to quarks are given by:
\begin{equation}
\xi_u = \frac{\cos{\alpha}}{\sin{\beta}},\ \ \
\xi_d = - \frac{\sin{\alpha}}{\cos{\beta}}.
\end{equation}
The coupling of the leptons differs between the Type II and Flipped models, and are: 
\begin{equation}
\xi_\ell = \xi_d\ (\textrm{Type II})\, ,
\ \ \ \
\xi_\ell = \xi_u\ (\textrm{Flipped})\, .
\end{equation}
The limit where the couplings become same as the SM couplings is $\xi_u=\xi_d=\xi_\ell=1$.
Taking the ratio of the up-type and down-type quark couplings we get 
\begin{equation}
-\frac{\xi_u}{\xi_d} = \frac{1}{\tan{\alpha} \tan{\beta}}
=
\frac{\cos{(\beta - \alpha)} + \cos{(\beta + \alpha)}}{
	\cos{(\beta - \alpha)} - \cos{(\beta + \alpha)}}.
\end{equation}
Now $|\xi_u/\xi_d| \sim 1$ for two cases i.e. either  $\beta - \alpha = \pi/2$ or $\beta + \alpha = \pi/2$:
\begin{itemize}
 \item $\beta - \alpha = \pi/2$ implies $\xi_d=+1$ which is called the \textbf{right-sign solution}.
 \item $\beta + \alpha = \pi/2$ implies $\xi_d=-1$ which is the\textbf{ wrong-sign solution}.
\end{itemize}
Moreover, one can further simplify things by noting that
\begin{equation}
\frac{\sin{(\beta - \alpha)}}{\sin{(\beta + \alpha)}}
= \frac{1-\frac{t_\alpha}{t_\beta}}{1+\frac{t_\alpha}{t_\beta}}
= \frac{1+\frac{1}{t_\beta^2} \frac{\xi_d}{\xi_u}}{1-\frac{1}{t_\beta^2} \frac{\xi_d}{\xi_u}}.
\end{equation}
thus, for large values of $t_\beta$ and $|\xi_u/\xi_d| \sim 1$ one can write
\begin{equation}
\sin{(\beta - \alpha)}
\sim
\sin{(\beta + \alpha)}
\left(
1 + \frac{2}{t_\beta^2} \frac{\xi_d}{\xi_u}
\right).
\end{equation}
For somewhat large values of $\tan{\beta}$, $ \sin{(\beta - \alpha)}\sim\sin{(\beta + \alpha)} $. Thus, for all values of $\tan{\beta}$, the SM limit corresponds to $\sin{(\beta-\alpha)} \sim 1$ ($\xi_d=1$), and for large values of $\tan{\beta}$, it also corresponds to $\sin{(\beta+\alpha)} \sim 1$ ($\xi_d=-1$). Therefore, even when the quark couplings are of the same magnitude as the SM values, one can have a wrong-sign solution where the magnitude of the couplings will be SM like but the overall sign will be opposite. Such wrong-sign solutions cannot be constrained from just measuring the $h \to b \bar{b}$ decay more precisely.

However, the situation is very different for $h \to \Upsilon \gamma$ decay. The wrong-sign solution modifies the direct diagram of $h \to \Upsilon \gamma$ by changing the overall sign of the amplitude of the process. Hence, the previously discussed near complete cancellation between direct and indirect diagram in SM will now be incomplete. 
\par It should be noted that although we have so far discussed the possible modifications and wrong-sign solution of the $h b \bar{b}$ coupling in the context of 2HDMs as  prototype models, the discussion applies to all models with a modified $h b \bar{b}$ coupling. In any BSM model, the modification to the $h \to \Upsilon \gamma$ decay width due to any variation in the $h b \bar{b}$ coupling can be calculated as:
\begin{align}
\Gamma[h \to \Upsilon \gamma]
&=\frac{1}{8 \pi} \frac{m_h^2 - m_\Upsilon^2}{m_h^2}
\left| \xi_b {\cal A}_\textrm{direct} + {\cal A}_\textrm{indirect} \right|^2 ,
\end{align}
where the factor $\xi_b$ takes into account any deviation, of magnitude or sign, from the SM value of the $h b \bar{b}$ coupling.
In particular for the case of wrong-sign solution with $|\xi_b| = 1$ we have
\begin{align}
\Gamma[h \to \Upsilon \gamma] &=\frac{1}{8 \pi} \frac{m_h^2 - m_\Upsilon^2}{m_h^2}
\left| - {\cal A}_\textrm{direct} + {\cal A}_\textrm{indirect} \right|^2 .
\label{eq:direc-wrong-sign}
\end{align}
Note that in \eqref{eq:direc-wrong-sign}, if there is a sign change in the contribution from the direct diagram ($ \xi_b<0 $), there will be constructive interference between the direct and indirect diagrams instead of destructive interference like in the SM.

\begin{figure}[!htbp]
	\centering
	\includegraphics[width=0.8\linewidth]{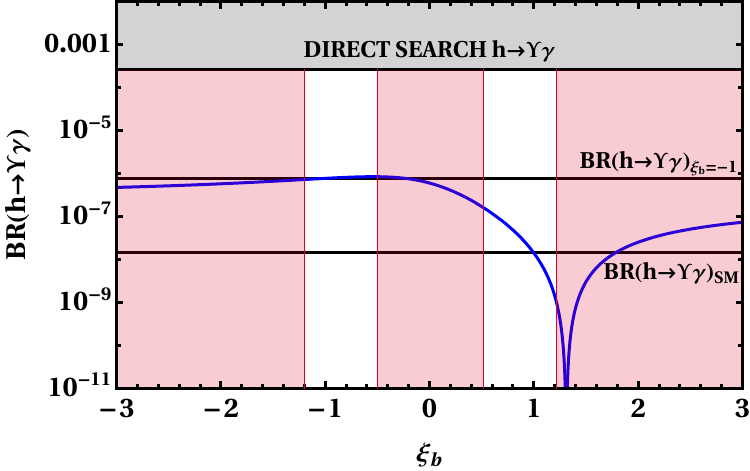}
	\caption{Branching Ratio of $h \to \Upsilon \gamma$ as a function of $\xi_b$ with $\xi_b = 1$ corresponding to the SM case.  The black lines show the wrong-sign and right-sign predictions for $|\xi_b| = 1$. The horizontal grey region depicts the region excluded by direct search of $ h \to \Upsilon \gamma $~\cite{ATLAS:2022rej} while the red regions are ruled out from $h \to X X$ measurements listed in Table \ref{tab:mu}~\cite{CMS-PAS-HIG-21-018,ATLAS-CONF-2025-006}.}
	\label{fig:UpGkD}
\end{figure}

The modification in the $h \to \Upsilon \gamma$ decay width due to either a magnitude or a sign change in the value of the $h b \bar{b}$ coupling $\xi_b$ is shown in Fig. \ref{fig:UpGkD}.
Fig. \ref{fig:UpGkD} shows the effect on $h \to \Upsilon \gamma$ branching ratio due to modifications in the $h b \bar{b}$ coupling parametrized by the $\xi_b$ parameter. As one can see, large modifications from the SM value in the magnitude of the $h b \bar{b}$ coupling are ruled out by the various  $h \to X X$ measurements at the LHC~\cite{CMS-PAS-HIG-21-018,ATLAS-CONF-2025-006}, where $X$ represents a possible decay product of $h$. The gray shaded region is excluded by the LHC constraint $\text{BR}(h\to\Upsilon\gamma) < 2.5 \times 10^{-4}$~\cite{ATLAS:2022rej}. Thus, a large change in the $h \to \Upsilon \gamma$ branching ratio due to a large change in the magnitude of the $h b \bar{b} $ coupling is not possible. This leaves us with a wrong-sign solution with $|\xi_b| \sim 1$ as the only possibility which can result in an order of magnitude change in the $h \to \Upsilon \gamma$ branching ratio.

As shown in Figure \ref{fig:UpGkD}, the right-sign solution $\xi_b = 1$ leads to a very small $h \to \Upsilon \gamma$ branching ratio\footnote{This value is highly sensitive to the input parameters. The value $1.41 \times 10^{-8}$ is taken from Ref.~\cite{Bodwin:2013gca} while our own computation gives a slightly different value of $9.70 \times 10^{-9}$.},
\begin{equation}
9.70 \times 10^{-9} \lsim \text{BR}(h \to \Upsilon \gamma) \lsim 1.41 \times 10^{-8}
\label{eq:SMvalue}
\end{equation}
whereas the wrong-sign solution $\xi_b = - 1$ leads to a branching ratio of
\begin{equation}
\text{BR}(h \to \Upsilon \gamma) = 7.66 \times 10^{-7}
\label{eq:wrong}
\end{equation}
which is larger by almost two orders of magnitude. Thus, we can conclude with very high confidence that the  wrong-sign solution $\xi_b \approx - 1$ is the only possibility that appreciably modifies the  $h \to \Upsilon \gamma$ branching ratio through changes in the direct diagram. 

\section{Possible modifications to the indirect diagram}
\label{sec:indirect}

\begin{figure}[!htbp]
	\centering
	\includegraphics{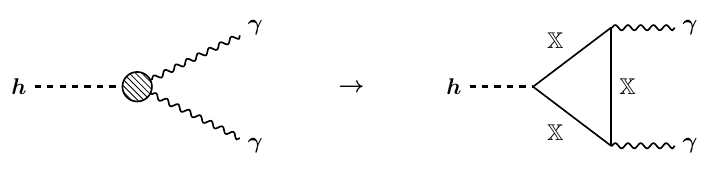}
	\caption{Feynman diagram for the $h \to \gamma \gamma$ process with a new particle $ \mathbb{X} $ running in the loop.}
	\label{fig:indirect-loop}
\end{figure}

\begin{figure}[!htbp]
	\centering
	\includegraphics[width = 0.7\textwidth]{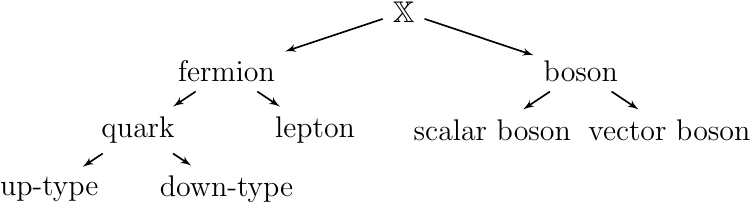}
	\caption{Possible particle types for $ \mathbb{X} $.}
	\label{fig:possible-X}
\end{figure}

Having looked at all possible modifications in the direct diagram, now let us look at the possible modifications in the indirect diagram. The only way to significantly modify the contribution of the indirect  decay diagram of $h \to \Upsilon \gamma$ is through possible modifications in the $h \to \gamma \gamma$ effective coupling. As shown schematically in Fig. \ref{fig:indirect-loop}, this can happen if any new particle runs in the $h \to \gamma \gamma$ loop. 
Here $ \mathbb{X} $ can be any electromagnetically charged particle so that the $ \mathbb{X}\mathbb{X} \, \gamma $ vertex exists. Thus, their are many possibilities as pictorially depicted in Fig. \ref{fig:possible-X}.

The modification to $h \to \Upsilon \gamma$ decay width due to variation in the $h \to \gamma \gamma$ decay coming from the contribution of new particles to the indirect diagram can be parametrized as:
\small\begin{equation}
\Gamma[h \to \Upsilon \gamma]
= \frac{m_h^2 - m_\Upsilon^2}{8\pi\,m_h^2}
\left| {\cal A}_\textrm{direct} + \sqrt{1+\frac{\text{err}}{100}}{\cal A}_\textrm{indirect} \right|^2 ,
\end{equation}
\normalsize
where 
\begin{equation}
\text{err} = \frac{\Gamma_{\rm new}[h \to\gamma\gamma] - \Gamma_{\rm SM}[h\rightarrow\gamma\gamma]}{\Gamma_{\rm SM}[h \to\gamma\gamma]} \times 100,
\end{equation}
is the variation (in \%) to the $ h \to\gamma\gamma $ decay width due to new physics.

To estimate the modification induced by all these particles in the indirect decay diagram one has to look at each possibility one by one.  To calculate the modifications to the Higgs decay width and production cross-section, new model files were written in SARAH~\cite{Staub:2015kfa} and FeynRules~\cite{Alloul:2013bka}, the mass spectrum and effective vertices were calculated in SPheno~\cite{Porod:2003um,Porod:2011nf}, and collider simulations were done in MadGraph~\cite{Alwall:2014hca}.

Among the various possibilities shown in Fig.~\ref{fig:possible-X}, the fermion loops actually decrease the $\text{BR}(h \to \Upsilon \gamma)$.
This is because the SM contribution is dominated by the  $ W $ boson loop and the contribution of a new fermion will be opposite in sign to the bosonic loop leading to destructive interference. 
New charged bosons interfere constructively with the existing SM $h\to\gamma\gamma $ loops since their contribution has the same sign as the SM bosonic loop. Among the bosons, we found that a new charged vector boson leads to the largest positive change in the $ h\to\Upsilon\gamma $ decay width, which we discuss below. All other cases are given in detail in Appendix~\ref{sec:indirect_appendix}.

\subsection{New  Heavy Charged Vector Bosons}
\label{sec:newVB}
We add a new heavy charged vector boson $ W'^+ $ to the SM particle content. We look at the simplest model-independent possibility of a new $SU(2)_L$ singlet vector boson without any connection to a gauge group. The conclusions drawn can be generalized to other cases. The charges of the new charged vector boson are given in Table \ref{tab:vec-bos}.
\begin{table}[!htbp]
	\begin{center}
		\begin{tabular}{|c|c|c|c|}
			\hline
\hspace{0.1cm}Name \hspace{0.1cm}& \hspace{0.1cm}Spin \hspace{0.1cm} & \hspace{0.1cm} Generations \hspace{0.1cm}& \hspace{0.1cm}\((\text{SU}(3)_C\otimes\, \text{SU}(2)_L\otimes\, U(1)_Y)\) \hspace{0.1cm}\\ 
			\hline  
			\(W'^+\) & 1  & 1 & \(({\bf 1},{\bf 1},1) \) \\ 
			\hline
		\end{tabular} 
	\end{center}
	\caption{Charges of the new charged vector boson.}
	\label{tab:vec-bos}
\end{table}
The relevant new terms in the Lagrangian in this case are given by:
\begin{align}
\mathcal{L} &= - \frac{1}{4} W'^+_{\mu\nu} W'^{\mu\nu -} + m_{W'}^2 W'^-_\mu W'^{+ \mu} + g_{hW'} h W'^-_\mu W'^{+ \mu} \\&+ 
 \frac{e}{\sin{\theta_W}} \left[
\frac{g_L}{\sqrt{2}}
\left(
\bar{\nu}_L \gamma^\mu C_L^\ell\, e_L
+
\bar{u}_L \gamma^\mu C_L^q\, d_L
\right) W'^+_\mu
+
\frac{g_R}{\sqrt{2}}
\left(
\bar{u}_R \gamma^\mu C_R^q\, d_R
\right) W'^+_\mu + \text{h.c.}
\right]
 \nonumber \,
\end{align}
where $ W'^+_{\mu\nu} = (\partial_\mu W'^+_\nu - \partial_\nu W'^+_\mu) $, $ h $ is the SM Higgs boson, and $\nu_L$, $e_{L,R}$, $u_{L,R}$, and $d_{L,R}$ denote the SM neutrinos, left- and right-handed charged leptons, up-type quarks, and down-type quarks, respectively. For the sake of simplicity, we take $C_L^{q,\ell}$ and $C_R^q$ to be identity matrices, $g_L=0$, and $g_R=1$ such that $W'^+$ does not interfere with SM weak processes, thereby avoiding the consequent experimental bounds.

\begin{figure}[!htbp]
	\centering
	\includegraphics{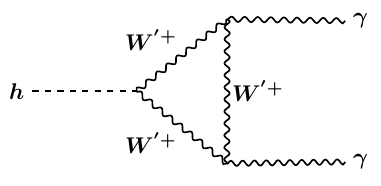}
    \hspace{0.5cm}
    \includegraphics{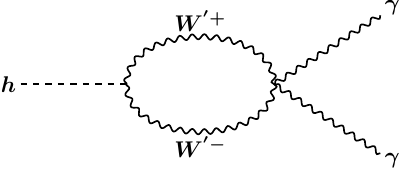}
	\caption{Feynman diagrams for the $h \to \gamma \gamma$ process with the new charged vector boson running in the loop.}
	\label{fig:htogamW}
\end{figure}

The Feynman diagrams for the process that contributes to the indirect diagram in $h \to \Upsilon \gamma$ decay in this case are shown in Fig. \ref{fig:htogamW}.
The contribution to the $h \to \gamma \gamma$ decay amplitude in this case can be calculated from the formula given in appendix \ref{appendix:contribution}:
\begin{equation}
{\cal A}_{new}^{W'^\pm}(h \to\gamma\gamma) = 
-g_{hV}\frac{v}{2 m_{W'}^2} A_1^{h} (\tau_{W'}) 
\label{Wamp}
\end{equation}
where $ m_{W'} $ is the mass of the new charged vector boson, $\tau_{W'}=m^2_{h}/4m_{W'}^2$ and $g_{hV}=v\, g_{hW'}$. The form factor $A_{1}^h$ is given in Eq. \eqref{eq:Ascalar}.
\par
\begin{figure}[!htbp]
	\centering
	\includegraphics[width=0.5\linewidth]{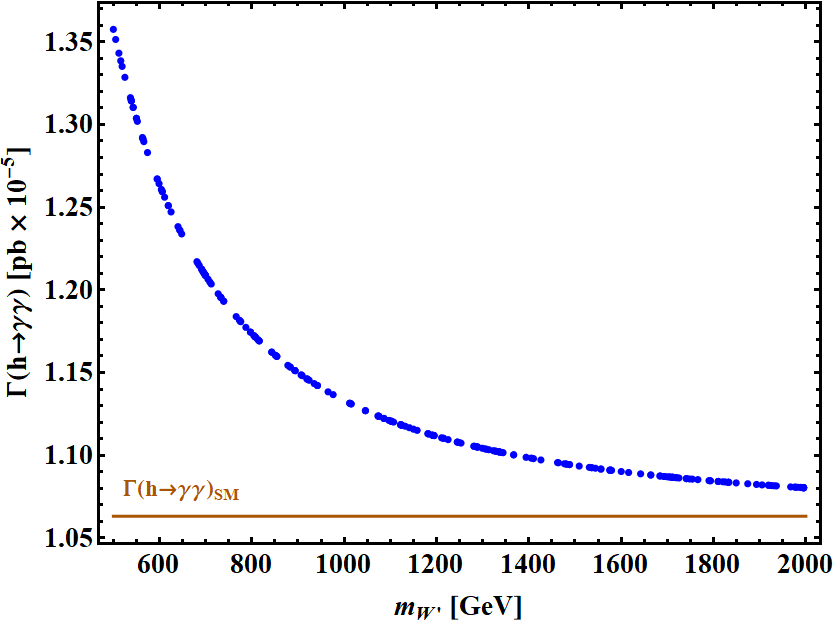}
	\caption{$ h \to \gamma \gamma $ decay width vs mass of the new charged vector boson for $ g_{hW'}=1 $.}
	\label{fig:decayWboson}
\end{figure}

 The modification in the $h \to \gamma \gamma$ decay width can be seen in Fig. \ref{fig:decayWboson}.
As we can see in Fig. \ref{fig:decayWboson}, the new charged vector boson interferes constructively with the existing SM $ h\to\gamma\gamma $ loops since its contribution has the same sign as the SM bosonic loop. 
\par
\begin{figure}[!htbp]
	\centering
	\includegraphics{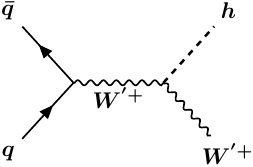}
	\caption{Feynman diagram for $ q\bar{q} \to W'^+ + h $ process involving the new charged vector boson.}
	\label{fig:prodW}
\end{figure}

\begin{figure}[!htbp]
	\centering
	\includegraphics[width=0.5\linewidth]{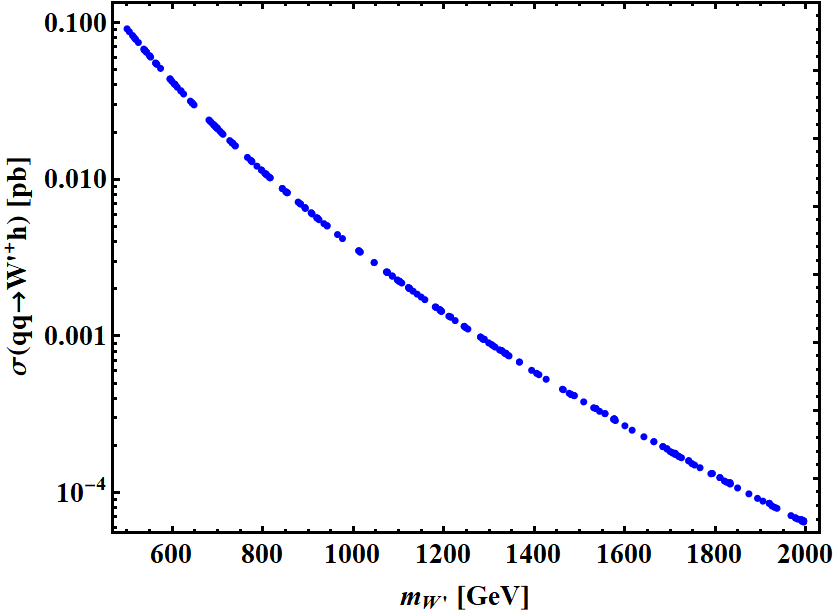}
	\caption{$ p p \to h $ production cross-section vs mass of the new charged vector boson for $ g_{hW'}=1 $.}
	\label{fig:prodplotW}
\end{figure}

\begin{figure}[!htbp]
	\centering
	\includegraphics[width=0.475\linewidth]{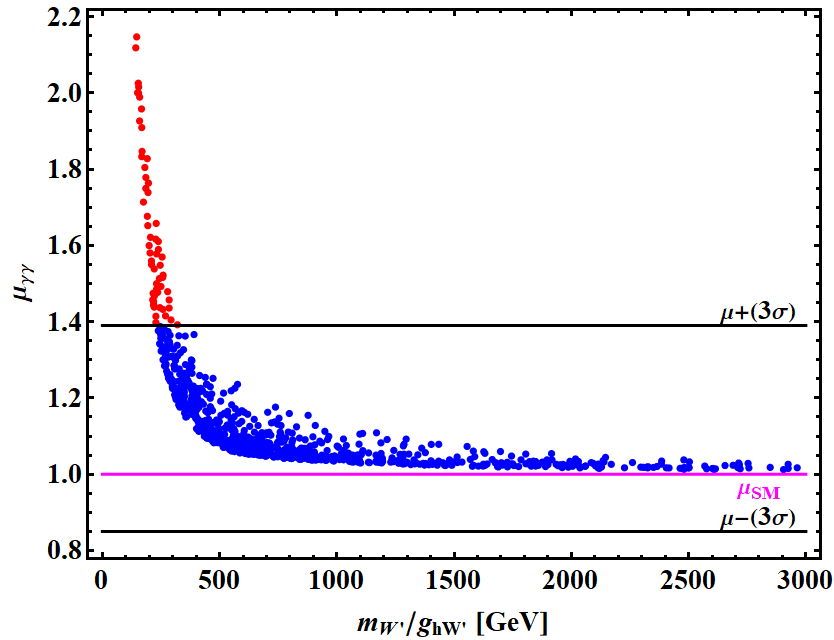}
	\includegraphics[width=0.515\linewidth]{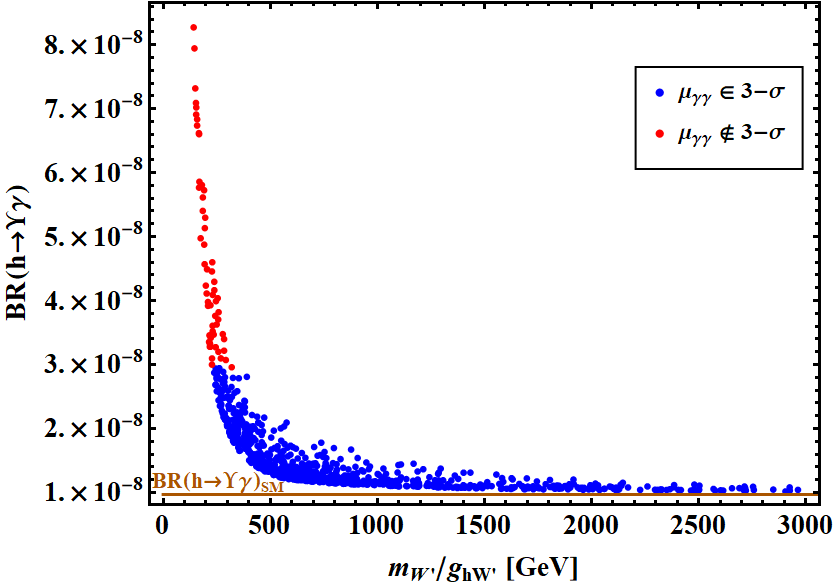}
	\caption{Left: $ \mu_{\gamma \gamma } $ vs ratio of mass of the new charged vector boson to $ g_{hW'} $. The solid black lines are the experimentally measured $ \mu_{\gamma \gamma } $ at $\pm3\sigma$ uncertainty \cite{CMS-PAS-HIG-21-018}. The magenta line is $ \mu_{\gamma \gamma } $ in the SM. 
	Right: Branching Ratio of $h \to \Upsilon \gamma$ vs ratio of mass of the new charged vector boson to $ g_{hW'} $ depicting the allowed values of $ \mu_{\gamma \gamma } $ at each point. }	
	\label{fig:muBRW}
\end{figure}

 Furthermore, there is a modification to the production of $ h $ in this case since the existence of the new charged vector boson modifies the $ q\bar{q} \to V + h $ and $ qq \to qq + (V^* V^* \to h) $ processes. The most constraining production process out of these is $ q\bar{q} \to W'^+ + h $. The Feynman diagram corresponding to this process is shown in Fig. \ref{fig:prodW}.
The modification to the production of $ h $ is shown in Fig. \ref{fig:prodplotW}.
As we can see in Fig. \ref{fig:prodplotW}, there is an increase in the production of $ h $ because of the existence of new production processes.
The overall modification to the $ \mu $-parameter and the $h \to \Upsilon \gamma$ branching ratio as a result is shown in Fig. \ref{fig:muBRW}.
The points in Fig. \ref{fig:muBRW} are generated assuming the coupling $ g_{hW'} $ to be in the perturbative range ($ \le\sqrt{4\pi} $). Since the new charged vector boson interferes constructively with the existing SM $ h\to\gamma\gamma $ loops,  the $ h\to\Upsilon\gamma $ decay width is also increased. Thus this case can lead to an enhancement in the $ h\to\Upsilon\gamma $ branching ratio. Again, the 3-$\sigma$ constraints from $h\to\gamma\gamma$ limits it to be only a small enhancement as shown in Fig. \ref{fig:muBRW}.

\subsection{Summarizing the contribution of new particles to the indirect diagram}
The variation to $ \Gamma[h \to\gamma\gamma] $ that can bring about the largest positive change in the $h \to \Upsilon \gamma$ decay width within experimental constraints on $ \mu_{\gamma\gamma} $ observed in the calculations is for the new charged vector boson at mass $ 753 $ GeV and coupling $ g_{hW'} = 3.07 $.
For these mass and coupling values,
\begin{equation}
\text{err} = \frac{\Gamma_{new}^{W'^+}[h \to\gamma\gamma] - \Gamma_{SM}[h \to\gamma\gamma]}{\Gamma_{SM}[h \to\gamma\gamma]} \times 100 = 38.24\% \, ,
\label{eq:Werr}
\end{equation}
for which the $h \to \Upsilon \gamma$ branching ratio is
\begin{equation}
\label{eq:indirectmod}
\text{BR}(h \to \Upsilon \gamma) = 2.94 \times 10^{-8},
\end{equation}
which is still smaller by more than one order of magnitude than the branching ratio due to the wrong-sign solution (from Eq.~\eqref{eq:wrong}),
\begin{equation}
\text{BR}(h \to \Upsilon \gamma) = 7.66 \times
10^{-7}.
\end{equation}
\par The variation to $ \Gamma(h \to\gamma\gamma) $ that can bring about the largest negative change in the $h \to \Upsilon \gamma$ decay width within experimental constraints on $ \mu_{\gamma\gamma} $ observed in the calculations in the sections above is in the case of a new up-type quark at mass $ 753 $ GeV and coupling $ g_{hT} = 1.91 $.
For these mass and coupling values,
\begin{equation}
\text{err} = \frac{\Gamma_{new}^{T'}[h \to\gamma\gamma] - \Gamma_{SM}[h \to\gamma\gamma]}{\Gamma_{SM}[h \to\gamma\gamma]} \times 100 = -18.62\%\, ,
\label{eq:uperr}
\end{equation}
for which the $h \to \Upsilon \gamma$ branching ratio is
\begin{equation}
\label{eq:indirectmod}
\text{BR}(h \to \Upsilon \gamma) = 3.18 \times 10^{-9}.
\end{equation}
The branching ratio in the SM is (from eq.~\eqref{eq:SMvalue})
\begin{equation}
\text{BR}(h \to \Upsilon \gamma) = 9.70 \times 10^{-9}.
\end{equation}
Thus, the SM value is larger by almost an order of magnitude.
\begin{figure}[!htbp]
	\centering
	\includegraphics[width=0.8\linewidth]{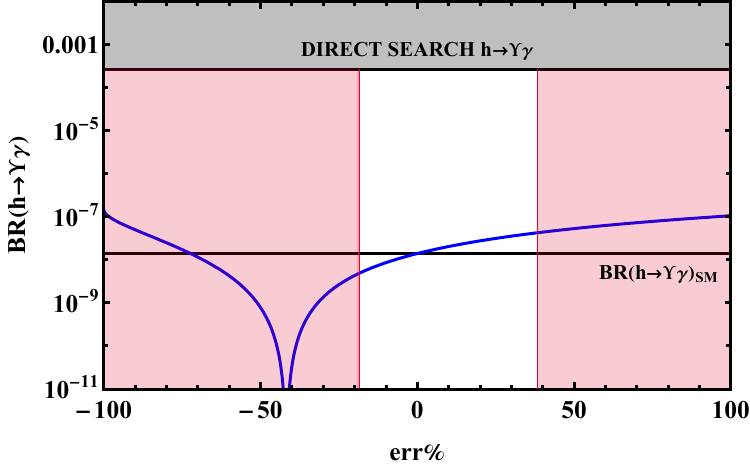}
	\caption{Branching Ratio of $h \to \Upsilon \gamma$ as a function of $ \text{err}\% $ in the $h \to \gamma \gamma$ decay width with $\text{err} = 0$ corresponding to the SM case. The black line shows the SM prediction. The grey region depicts the region excluded by direct search of $ h \to \Upsilon \gamma $~\cite{ATLAS:2022rej} while the red regions are ruled out from constraints on $ \mu_{\gamma \gamma } $ (Eqs. \eqref{eq:Werr} and \eqref{eq:uperr}).}
	\label{fig:UpGerr}
\end{figure}
\par The modification in the $h \to \Upsilon \gamma$ decay width due to variation in the $h \to \gamma \gamma$ decay can be seen in Figure \ref{fig:UpGerr}. The left red vertical band in Fig. \ref{fig:UpGerr} shows the minimum allowed value of  $ \text{err} $ which is seen in the case of the new up-type quark. The region to the left of this band is hence ruled out by constraints on $ \mu_{\gamma \gamma } $. The right red vertical band shows the maximum allowed value of  $ \text{err} $ which is seen in the case of the new charged vector boson. The region to the right of this band is hence ruled out by constraints on $ \mu_{\gamma \gamma } $. Hence, we can conclude that the modification to the indirect diagram coming from any type of new physics is not enough to enhance the $h\to\Upsilon\gamma$ branching ratio by two orders of magnitude compared to the SM prediction.\\
\par Thus, the wrong-sign solution is the only possibility that gives rise to a roughly two orders of magnitude larger $h \to \Upsilon \gamma$ decay width.
Although the present bound on $h \to \Upsilon \gamma$ is relatively weak, its future prospects are promising. Among future colliders, the expected number of events for the SM and wrong-sign cases are  $N_{\rm event}^{\rm SM}=0.7$, $N_{\rm event}^{\rm wrong-sign}=114$ at the HL-LHC with luminosity $3 \text{ ab}^{-1}$. For the FCC-hh with luminosity $1 \text{ ab}^{-1}$ we have $N_{\rm event}^{\rm SM}=3.6$, and $N_{\rm event}^{\rm wrong-sign}=608$. They thus have potential to probe the wrong-sign solution. Furthermore, in the future it might be possible to further increase the signal significance and thus the discovery potential by adopting advanced Machine learning~(ML) based analysis techniques~\cite{Gambhir:2025afb} at the HL-LHC or FCC-hh.

\vspace{1cm}
\section{Conclusions}
\label{sec:conclusions}

In conclusion, the rare Higgs decay mode $h \to \Upsilon \gamma$ provides us with a unique testing ground for new physics beyond the SM. This is owing to the fact that in the SM, the direct and indirect Feynman diagrams interfere destructively, almost completely canceling each other. This accidental cancellation makes the SM decay width very small. Thus, any new physics which can disrupt this accidental cancellation can potentially modify the $h \to \Upsilon \gamma$ decay width by orders of magnitude. Furthermore, due to presence of interference terms, even the information about the sign of the new physics couplings, typically lost in most other decay modes like $h \to b \bar{b}$, can be probed in this decay channel. 

In this work we have carried out a systematic and model independent analysis of possible modifications to the $h \to \Upsilon \gamma$ decay coming from new physics. Taking into account the bounds on Higgs production and decay processes from experimental measurements at the LHC, possible modifications have been obtained in the $\Gamma[h \to \gamma \gamma]$ and hence $\Gamma[h \to \Upsilon \gamma]$ decay processes due to presence of various types of new particles. We found that while the accidental cancellation of the SM direct and indirect amplitudes can indeed be disrupted by presence of new physics and new particles, the most significant change in $\Gamma[h \to \Upsilon \gamma]$ occurs if the Higgs has a wrong-sign $h b \bar{b}$ coupling. 

In fact, due to a wrong-sign $h b \bar{b}$ coupling, the  $h \to \Upsilon \gamma$ decay width increases by almost two orders of magnitude compared to the SM value. We further found that no other new physics can change the SM value of the $h \to \Upsilon \gamma$ decay width by such a large amount. Thus, $h \to \Upsilon \gamma$ offers the most promising probe for wrong-sign $h b \bar{b}$ coupling, particularly due to the fact that direct $h b \bar{b}$ decay measurements cannot distinguish between right sign and wrong-sign $h b \bar{b}$ couplings.
Finally, if the LHC or any future collider ever observes the $h \to \Upsilon \gamma$ decay mode with a significantly enhanced decay width compared to the SM value, then it will be a conclusive evidence of wrong-sign $h b \bar{b}$ coupling and the presence of more than one Higgs boson.


\begin{acknowledgments}
The work of S.M. is supported by KIAS Individual Grants (PG086002) at Korea Institute for Advanced Study.
The work of R.S. is supported by the Government of India, SERB Startup Grant SRG/2020/002303.
The work of A.B. is supported by Fundação para a Ciência e a Tecnologia (FCT, Portugal) through the PhD grant UI/BD/154391/2023 and through the projects UID/00777/2025 (https://doi.org/10.54499/UID/00777/2025) and CERN/FIS-PAR/0019/2021.
\end{acknowledgments}

\appendix


\section{Measuring the couplings of the Higgs boson}
\label{sec:measure}

\begin{figure}[!htbp]
	\centering
	\begin{subfigure}{\textwidth}
		\includegraphics{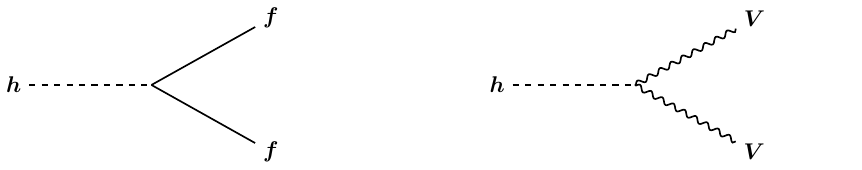}
		\caption{Tree-level Higgs boson decays in the $ h\rightarrow ff $ (left) and $ h\rightarrow VV $ (right) channels. Here $ f $ denotes fermions and $ V $ denotes the vector bosons $ W $ and $ Z $.  }
	\end{subfigure}
	\begin{subfigure}{\textwidth}
		\includegraphics{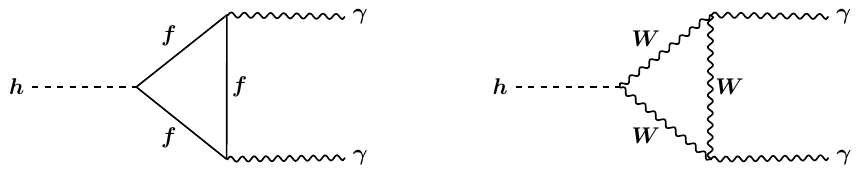}
		\caption{Loop-level Higgs boson decay in the $ h\rightarrow \gamma\gamma $ channel with $ f $ (left) and $ W $ (right) running in the loop. Here $ \gamma $ denotes photons. }
	\end{subfigure}
	\caption{Feynman diagrams for some leading order contributions to the SM Higgs boson decays. }
	\label{fig:feyn-h-to-xx}
\end{figure}

Here we present a brief discussion of the potential decay modes of the $125$ GeV scalar. For the sake of brevity, throughout this work we will simply call it the Higgs ($h$) boson\footnote{Note that we are not identifying it with the SM Higgs particle. Whenever we will discuss the SM case, we will call it the ``SM Higgs'' boson.}.
The Higgs boson couplings can be constrained by measuring the decay widths of it to various particles. As shown in Fig. \ref{fig:feyn-h-to-xx}, they can be broadly classified into two categories a) Higgs decay through its direct coupling with other particles and b) loop induced decays, the most prominent such decay being $h \to \gamma \gamma$.  \\

\begin{figure}[!htbp]
	\centering
	\includegraphics[width=0.8\linewidth]{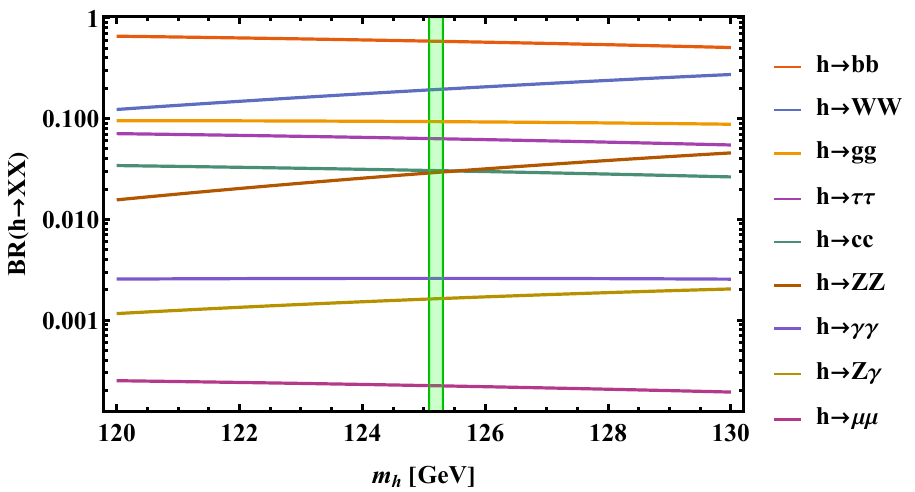}
	\caption{Branching ratios of various Higgs decay processes in the SM versus the mass of the Higgs boson. The experimentally measured mass is $ 125.20 \pm 0.11 $ GeV \cite{ParticleDataGroup:2024cfk}.}
	\label{fig:h-to-xx}
\end{figure}

The Higgs boson has several decay modes each quantified by the ratio called branching ratio given by 
\begin{eqnarray}
 BR_{XX}=\frac{\Gamma[h\rightarrow {XX}]}{\Gamma[h\rightarrow all]}
\end{eqnarray}
where $ \Gamma[h\rightarrow XX] $ is the decay width of the process $ h\rightarrow XX $ and $ \Gamma[h\rightarrow all] $ is the total decay width. The subscript $XX$ denotes that the branching ratio is for the decay mode $h \to X X$; $X$ being the given SM or BSM particle to which the Higgs is decaying. For the SM case, one can calculate the branching ratios theoretically and if the $125$ GeV scalar is the SM Higgs then they will be as shown in Fig. \ref{fig:h-to-xx}. 

Note that Fig. \ref{fig:h-to-xx} only shows the branching ratios for the first few leading decay modes. The Higgs has other rarer decay modes including decay modes involving hadrons. One such rare hadronic decay mode is $h \to \Upsilon \gamma$ on which we will concentrate in the later sections.

The signal strength modifier or $\mu$-parameter is often used to quantify Higgs boson decays in comparison to the SM predictions. It is defined as the ratio of the experimentally measured Higgs boson yield to the yield as predicted by the Standard Model. 
\begin{equation}
(\mu_{XX})_{\rm{Exp}} =  \frac{(\sigma)_{measured}}{(\sigma)_{SM}} \frac{(BR_{XX})_{measured}}{(BR_{XX})_{SM}}, \hspace{1cm}
\end{equation}
where $ \sigma $ is the Higgs production cross section and $ BR_{XX} $ is the branching ratio of a specific decay channel $ h \to X X $. 
To compare the calculations of new physics models with experimental measurements, the following formula is used to characterize the $ \mu $ parameter.  
\begin{equation}
(\mu_{XX})_{NP} = \frac{(\sigma)_{NP}}{(\sigma)_{SM}} \frac{(BR_{XX})_{NP}}{(BR_{XX})_{SM}}
\label{eq:mu}
\end{equation}
where $ (\sigma)_{NP} $ is the Higgs production cross section in the given new physics model  and $ (BR_{XX})_{NP} $ is the branching ratio of the decay channel $ h\rightarrow XX $ for the same new Physics model.

Since its discovery in 2012, both  CMS~\cite{CMS-PAS-HIG-21-018} and ATLAS~\cite{ATLAS-CONF-2025-006} have measured the $(\mu_{XX})_{\rm{Exp}}$ parameter for several Higgs decay modes which are summarized in Table~\ref{tab:mu}.

\begin{table}[!htbp]
	\centering
	\begin{tabular}{| c | c |}
		\hline
		\hspace{0.4cm}Decay mode \hspace{0.4cm}& \hspace{0.4cm}$(\mu_{XX})_{\rm{Exp}}$ \hspace{0.4cm} \\
		\hline
		&  \\[-10pt]
		
		$ h \to bb $  &  $1.02^{+0.16}_{-0.15}$  \\[5pt]
		
		$ h \to \tau\tau $  &  $0.84^{+0.10}_{-0.10}$  \\[5pt]
		
		$ h \to WW $  &  $0.99^{+0.09}_{-0.09}$  \\[5pt]
		
		$ h \to ZZ $  &  $1.03^{+0.11}_{-0.10}$  \\[5pt]
		
		$ h \to \gamma\gamma $  &  $1.12^{+0.09}_{-0.09}$  \\[5pt]
		
		$ h \to \mu\mu $  &  $1.15^{+0.44}_{-0.41}$   \\[5pt]
		\hline
	\end{tabular}
	\caption{Experimentally measured $ \mu $-parameters by CMS~\cite{CMS-PAS-HIG-21-018} collaboration with uncertainties ($\pm 1\sigma$) for various decay modes of $h$. The analogous ATLAS collaborations measurements can be found in~\cite{ATLAS-CONF-2025-006}.  }
	\label{tab:mu}
\end{table}	

As we can see in Table \ref{tab:mu}, due to background contamination and limitations in detection of final state particles, there is a considerable amount of uncertainty in the measurements of the Higgs boson decay modes. The decay to gauge bosons ($ ZZ $ and $ WW $) are relatively clean, resulting in lower uncertainties. The decay to quarks (such as $ b \bar{b} $) on the other hand, has a strong background from QCD multijet generation and thus a high amount of uncertainty. The decay to light charged leptons like the muon have a low background, but due to their small masses, they have small couplings with the Higgs boson, so their decay widths are relatively difficult to detect. The most significant experimental uncertainties in measuring the rate of decay to two photons ($ \gamma\gamma $) are related to photon identification.

Apart from the decay modes listed in Table \ref{tab:mu}, there are also several predicted SM Higgs decay processes with very small decay widths, which make them very hard to detect. These are called rare Higgs decays. Some of these processes are given in Table \ref{tab:rare} with their SM predicted branching ratios.

\begin{table}[!htbp]
	\centering
	\begin{tabular}{|c|c|}
		\hline
		\hspace{0.2cm}Decay mode \hspace{0.2cm}& \hspace{0.2cm}Branching Ratio \hspace{0.2cm}\\
		\hline
		& \\[-10pt]
		
		$ h \to \rho^0\gamma $  &  $ 1.68 \times 10^{-5} $ \\[5pt]
		
		$ h \to \omega\gamma $  & $ 1.48 \times 10^{-6} $ \\[5pt]
		
		$ h \to \phi\gamma $  & $ 2.31 \times 10^{-6} $ \\[5pt]
		
		$ h \to J/\psi\gamma $  & $ 2.46 \times 10^{-6} $ \\[5pt]
		
		$ h \to \Upsilon (1S) \gamma $  & $ 1.41 \times 10^{-8} $ \\[5pt]
		\hline
	\end{tabular}
	\caption{Branching ratios of rare Higgs decay modes as predicted by the SM \cite{Konig:2015qat,Bodwin:2013gca}. Here, $ \rho^0 $, $ \omega $, $ \phi $, $ J/\psi $ and $ \Upsilon $ are mesons.}
	\label{tab:rare}
\end{table}	

Surprisingly, the $ h \to \Upsilon\gamma $ branching ratio is the smallest among the processes given in Table \ref{tab:rare} in-spite of the $ \Upsilon $ meson being the heaviest and thus having the largest coupling with the Higgs boson. This happens due to an accidental cancellation between the two diagrams involved in this process. 
Note that due to the close nature of the cancellation, the SM branching ratio of $h \to \Upsilon \gamma$ is extremely sensitive to the values of the input parameters and different groups have reported slightly different values of the SM branching ratios\cite{Bodwin:2013gca,Konig:2015qat,Zhou:2016sot,Brambilla:2019fmu}. However, since in our work we are interested in new physics effects which can change the branching ratio by orders of magnitude, the high sensitivity to input parameters will not have any appreciable effect on our results and conclusions. 

\section{Other possible modifications to the indirect diagram}
\label{sec:indirect_appendix}
In this section, we discuss one by one the modifications to the indirect decay diagram induced by all types of particles other than new heavy charged vector bosons (discussed in Sec.~\ref{sec:newVB}).

\subsection{New  Heavy Charged Leptons }

We begin with looking at the possibility of a new heavy vector-like charged lepton pair $l'_L, l'_R$. The charged lepton needs to couple with Higgs which depending on its $SU(2)_L$ charge can be through Yukawa coupling with SM charged leptons or through another new pair of charged leptons with appropriate $SU(2)_L$ charges. We look at the simplest possibilities of $SU(2)_L$ singlet and doublet leptons in details. The conclusions drawn can then be easily generalized to other cases.

\subsubsection{Vector-like singlet leptons}
\begin{figure}[!htbp]
	\centering
	\includegraphics{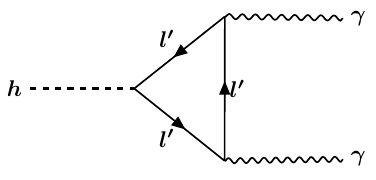}
	\caption{Feynman diagram for the $h \to \gamma \gamma$ process with the new lepton running in the loop.}
	\label{fig:htogamlep}
\end{figure}

The charges of the Vector-like singlet leptons are given in Table \ref{tab:singlep}.
\begin{table}[]
	\begin{center}	
		\begin{tabular}{|c|c|c|c|}
			\hline
\hspace{0.1cm} Name \hspace{0.1cm}& \hspace{0.1cm}Spin \hspace{0.1cm} & \hspace{0.1cm} Generations\hspace{0.1cm} &\hspace{0.1cm} \((\text{SU}(3)_C\otimes\, \text{SU}(2)_L\otimes\, U(1)_Y)\)\hspace{0.1cm} \\ 
			\hline  
			\(l'_L\) & \(\frac{1}{2}\)  & 1 & \(({\bf 1},{\bf 1},-1) \) \\ 
			\(l'_R\) & \(\frac{1}{2}\)  & 1 & \(({\bf 1},{\bf 1},-1) \) \\ 
			\hline
		\end{tabular} 
			\end{center}
			\caption{Charges of the new $SU(2)_L$ singlet leptons.}
			\label{tab:singlep}
\end{table}
The relevant new terms in the Lagrangian in this case are given by:
\begin{equation}
-\mathcal{L} = m_{l'} \bar{l'_{L}} l'_R + g_{hl} \bar{L}_i \Phi l'_R,
\end{equation}
where $ L_i = (\nu_{L,i},\, e_{L,i} )$; $i = 1,2,3$ are the SM lepton doublets and $ \Phi $ is the SM Higgs doublet. 
Since they have the same electric charge and spin, the new leptons mixes with the SM ones and the new mixed state acts as the 4th generation of the lepton family.
The Feynman diagram for the process that contributes to the indirect diagram in $h \to \Upsilon \gamma$ decay in this case is shown in Fig. \ref{fig:htogamlep}.
The contribution to the $h \to \gamma \gamma$ decay amplitude in this case can be calculated from the formula given in appendix \ref{appendix:contribution}:
\begin{equation}
{\cal A}_{new}^{l'}(h \to\gamma\gamma) = 
2 g_{hl} \frac{v}{m_{l'}} A_{1/2}^{h} 
(\tau_{l'}),
\label{lamp}
\end{equation}
where $ g_{hl}$ is the $h l' l'$ effective coupling, $ m_{l'} $ is the mass of the new lepton, $\tau_{l'}=m^2_{h}/4m_{l'}^2$ and $ v $ is the vacuum expectation value of the Higgs field. The form factor $A_{1/2}^h$ is given in Eq. \eqref{eq:Ascalar}.\\
\par
\begin{figure}[!htbp]
	\centering
\includegraphics[width=0.5\linewidth]{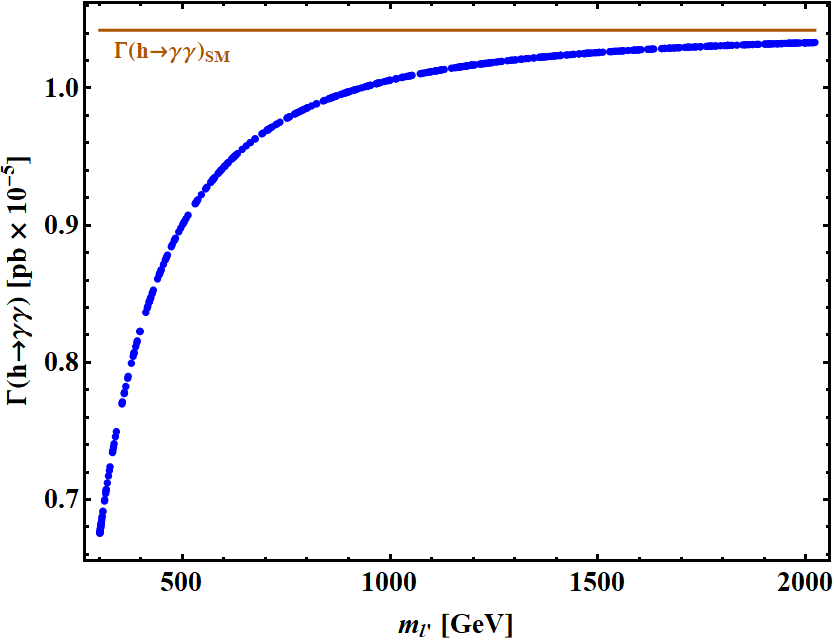}
	\caption{$ h \to \gamma \gamma $ decay width vs mass of the new lepton for $ g_{hl}=1 $.}
	\label{fig:lep-gaga-width}
\end{figure}

 The modification in the $h \to \gamma \gamma$ decay width is shown in Fig. \ref{fig:lep-gaga-width}.
As we can see in Fig. \ref{fig:lep-gaga-width}, the new lepton interferes destructively with the existing SM $ h\to\gamma\gamma $ loops. This is to be expected as the SM contribution is dominated by the  $ W $ boson loop and we have added a new fermion in the loop whose contribution will be opposite in sign to the bosonic loop leading to destructive interference~\footnote{Note that even in SM the fermionic loops (dominated by top loop) actually interfere destructively with the $W$ boson loop.}. 

\begin{figure}[!htbp]
	\centering
	\includegraphics[width=0.475\linewidth]{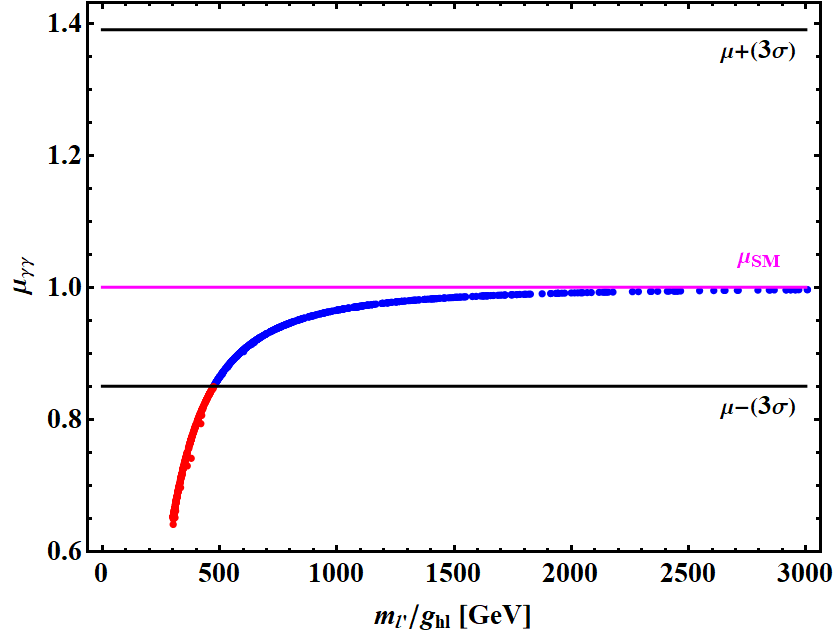}
	\includegraphics[width=0.515\linewidth]{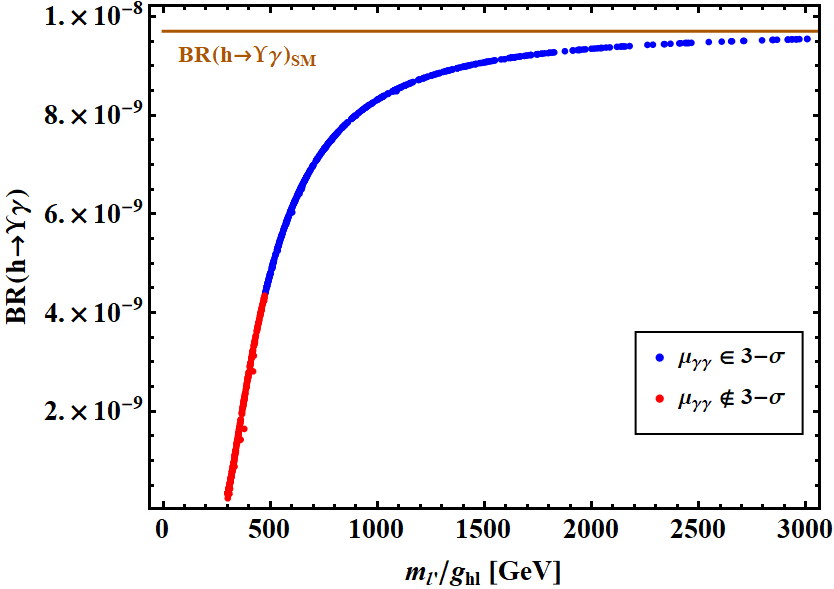}
	\caption{Left: $ \mu_{\gamma \gamma } $ vs ratio of mass of the new vector-like singlet lepton to $ g_{hl} $. The solid black lines are the experimentally measured $ \mu_{\gamma \gamma } $ at $\pm3\sigma$ uncertainty \cite{CMS-PAS-HIG-21-018}. The magenta line is $ \mu_{\gamma \gamma } $ in the SM. 
	Right: Branching Ratio of $h \to \Upsilon \gamma$ vs ratio of mass of the new vector-like singlet lepton to $ g_{hl} $ depicting the allowed values of $ \mu_{\gamma \gamma } $ at each point. }	
	\label{fig:cl-upga}
\end{figure}

Furthermore, there is no modification to the production of $ h $ in this case as the new lepton is not a coloured particle and hence, it does not run in the gluon fusion loop.
The overall modification to the $ \mu_{\gamma\gamma} $-parameter and the $h \to \Upsilon \gamma$ branching ratio are shown in Fig. \ref{fig:cl-upga}.
The points in Fig. \ref{fig:cl-upga} are generated assuming the coupling $ g_{hl} $ to be in the perturbative range ($ \le\sqrt{4\pi} $). Since the new lepton interferes destructively with the existing SM $ h\to\gamma\gamma $ loops,  the $ h\to\Upsilon\gamma $ decay width is also decreased. Thus this case cannot lead to any enhancement in the $ h\to\Upsilon\gamma $ branching ratio.

\subsubsection{Vector-like doublet leptons}

The charges of the Vector-like doublet leptons are given in Table \ref{tab:doublet-lep}.
\begin{table}[!htbp]
	\begin{center}
		\begin{tabular}{|c|c|c|c|}
			\hline
\hspace{0.1cm}Name \hspace{0.1cm}& \hspace{0.1cm}Spin \hspace{0.1cm} & \hspace{0.1cm} Generations \hspace{0.1cm}& \hspace{0.1cm}\((\text{SU}(3)_C\otimes\, \text{SU}(2)_L\otimes\, U(1)_Y)\) \hspace{0.1cm}\\ 
			\hline  
			\(l'_L\) & \(\frac{1}{2}\)  & 1 & \(({\bf 1},{\bf 2},-1) \) \\ 
			\(l'_R\) & \(\frac{1}{2}\)  & 1 & \(({\bf 1},{\bf 2},-1) \) \\ 
			\hline
		\end{tabular} 
	\end{center}
	\caption{Charges of the new $SU(2)_L$ doublet leptons.}
	\label{tab:doublet-lep}
\end{table}
The relevant new terms in the Lagrangian in this case are given by:
\begin{equation}
-\mathcal{L} = m_{l'} \bar{l'_{L}} l'_R + g_{hl} \bar{l'_{L}} \Phi e_R,
\end{equation}
where $ e_R $ are the SM right-handed charged leptons. Note that henceforth we will suppress the family generation index for the sake of brevity.
The  production and decay properties of the Higgs boson in this case are very similar to the case of vector-like singlet leptons. We again get a destructive interference leading to a decrease in  $ h\to\Upsilon\gamma $ branching ratio. Similar behaviour is observed for any $SU(2)_L$ representation of new charged leptons. Hence, we can conclude that additional charged leptons beyond SM cannot mimic the wrong-sign like enhancement of $ h\to\Upsilon\gamma $ branching ratio.

\subsection{New Heavy Quarks}

Now we move on to the case when new heavy quarks are added to the SM particle content. We start with the scenario where up-type quarks are added and will discuss the down-type quarks afterwards. Like the charged lepton case, for quarks also we can add particles belonging to different representations of $SU(2)_L$. As before we discuss the $SU(2)_L$ singlet cases first and will argue that the higher  $SU(2)_L$ representations will not change the conclusions. 

\subsubsection{Vector-like singlet up-type quarks}

\begin{figure}[!htbp]
	\centering
	\includegraphics{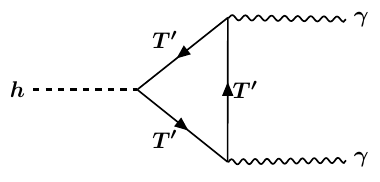}
	\caption{Feynman diagram for the $h \to \gamma \gamma$ process with the new up-type quark running in the loop.}
	\label{fig:upgdecay}
\end{figure}

\begin{figure}[!htbp]
	\centering
	\includegraphics[width=0.5\linewidth]{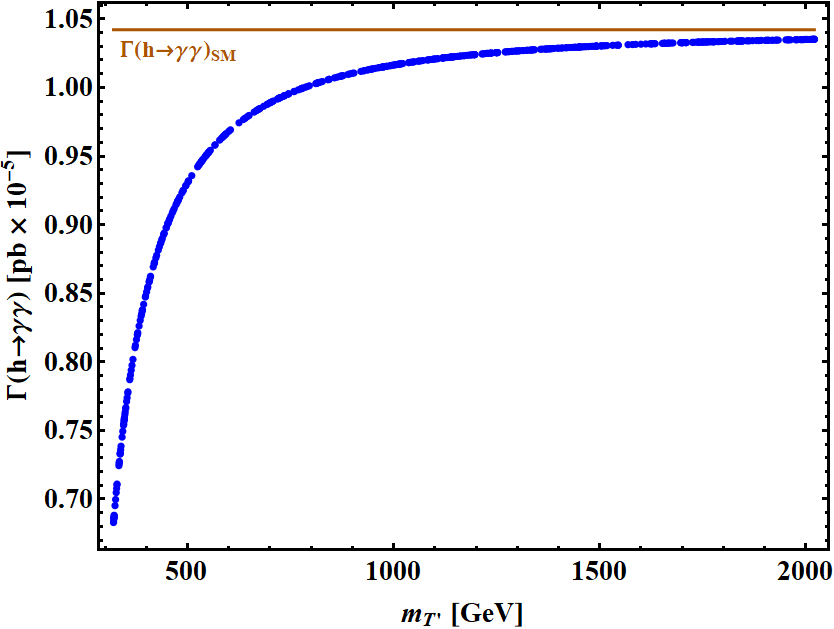}
	\caption{$ h \to \gamma \gamma $ decay width vs mass of the new up-type quark for $ g_{hT}=1 $.}
	\label{fig:uphgaga}
\end{figure}

We add a vector-like pair of heavy up-type quarks $T'_L, T'_R$ to the SM particle content with charges as shown in Table \ref{tab:singup}.
\begin{table}[!htbp]
	\begin{center}	
		\begin{tabular}{|c|c|c|c|}
			\hline
\hspace{0.1cm}Name \hspace{0.1cm}& \hspace{0.1cm}Spin \hspace{0.1cm} & \hspace{0.1cm} Generations\hspace{0.1cm} & \hspace{0.1cm}\((\text{SU}(3)_C\otimes\, \text{SU}(2)_L\otimes\, U(1)_Y)\) \hspace{0.1cm}\\ 
			\hline  
			\(T'_L\) & \(\frac{1}{2}\)  & 1 & \(({\bf 3},{\bf 1},2/3)\) \\ 
			\(T'_R\) & \(\frac{1}{2}\)  & 1 & \(({\bf 3},{\bf 1},2/3)\) \\ 
			\hline
		\end{tabular} 
	\end{center}
	\caption{Charges of the new $SU(2)_L$ singlet up-type quarks.}
	\label{tab:singup}
\end{table}
The relevant new terms in the Lagrangian in this case are given by:
\begin{equation}
-\mathcal{L} = m_{T'} \bar{T'_{L}} T'_R + g_{hT} \bar{Q} \tilde{\Phi} T'_R,
\end{equation}
where $ Q=(u_L,\,\, d_L) $ is the SM quark doublet and $ \tilde{\Phi} = i\sigma_2 \Phi^* $. $ \sigma_2 $ is the second Pauli matrix. 
Since they have the same electric charge and spin, the new up-type quark mixes with SM up-type quarks and the new mixed state acts as the 4th generation of the up-type quark family.
The Feynman diagram for the process that contributes to the indirect diagram in $h \to \Upsilon \gamma$ decay in this case is shown in Fig. \ref{fig:upgdecay}.
The contribution to the $h \to \gamma \gamma$ decay amplitude in this case can be calculated from the formula given in appendix \ref{appendix:contribution}:
\begin{equation}
{\cal A}_{new}^{T'}(h \to\gamma\gamma) = 
\frac{8}{3} g_{hT} \frac{v}{m_{T'}} A_{1/2}^{h} 
(\tau_{T'}),
\label{uamp}
\end{equation}
where $g_{hT}$ is the effective $h T' T'$ coupling,  $ m_{T'} $ is the mass of the new up-type quark and $\tau_{T'}=m^2_{h}/4m_{T'}^2$.
The modification in the $h \to \gamma \gamma$ decay width can be seen in Fig. \ref{fig:uphgaga}.
As we can see in Fig. \ref{fig:uphgaga}, the new up-type quark interferes destructively with the existing SM $ h\to\gamma\gamma $ loops since its contribution is opposite in sign to the SM bosonic loop (as in the case of a new lepton). 
\par
\begin{figure}[!htbp]
	\centering
	\includegraphics{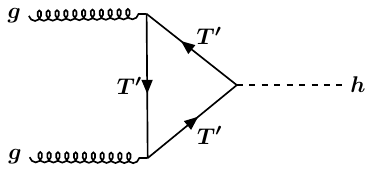}
	\caption{Feynman diagram for Higgs production via the gluon fusion process with the new up-type quark running in the loop.}
	\label{fig:gghup}
\end{figure}

\begin{figure}[!htbp]
	\centering
	\includegraphics[width=0.5\linewidth]{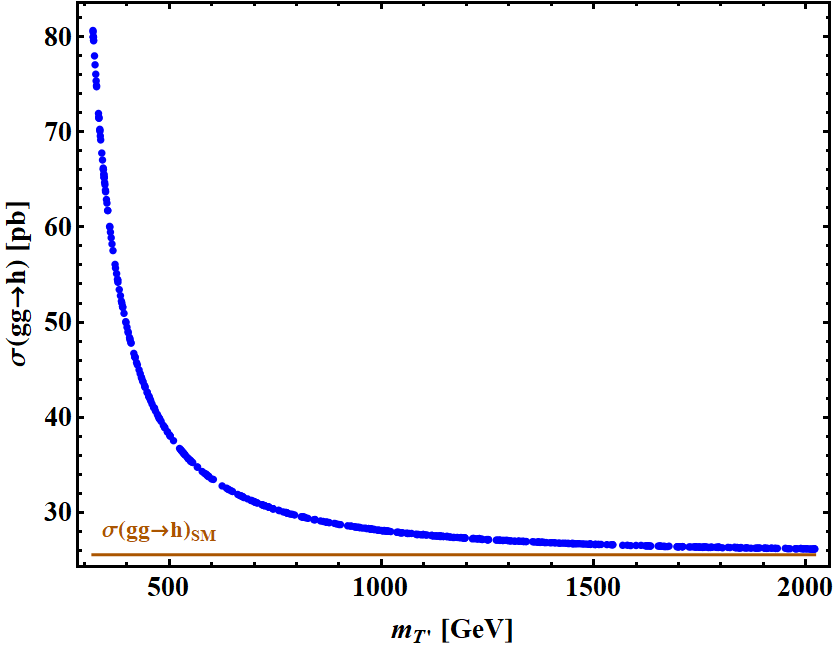}
	\caption{$ p p \to h $ production cross-section vs mass of the new up-type quark for $ g_{hT}=1 $.}
	\label{fig:upprod}
\end{figure}

\begin{figure}[!htbp]
	\centering
	\includegraphics[width=0.475\linewidth]{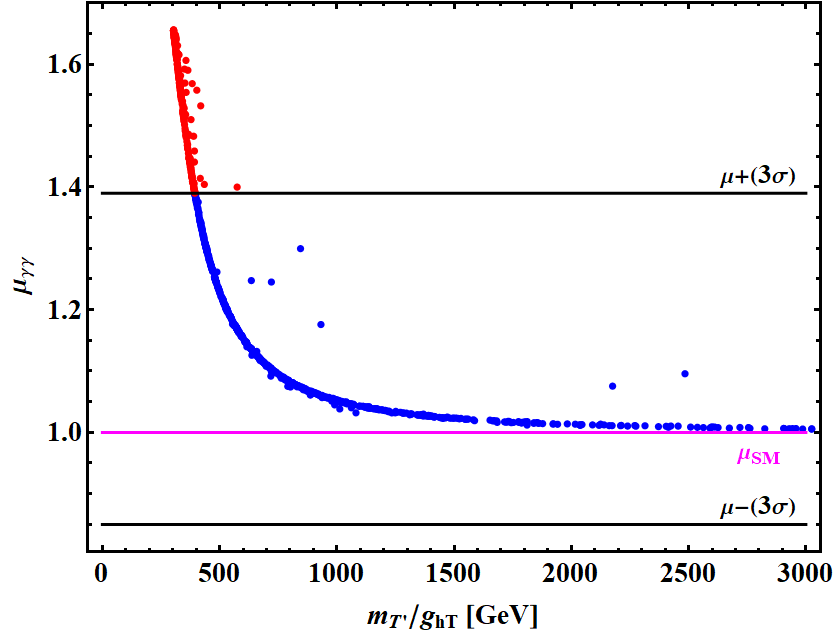}
	\includegraphics[width=0.515\linewidth]{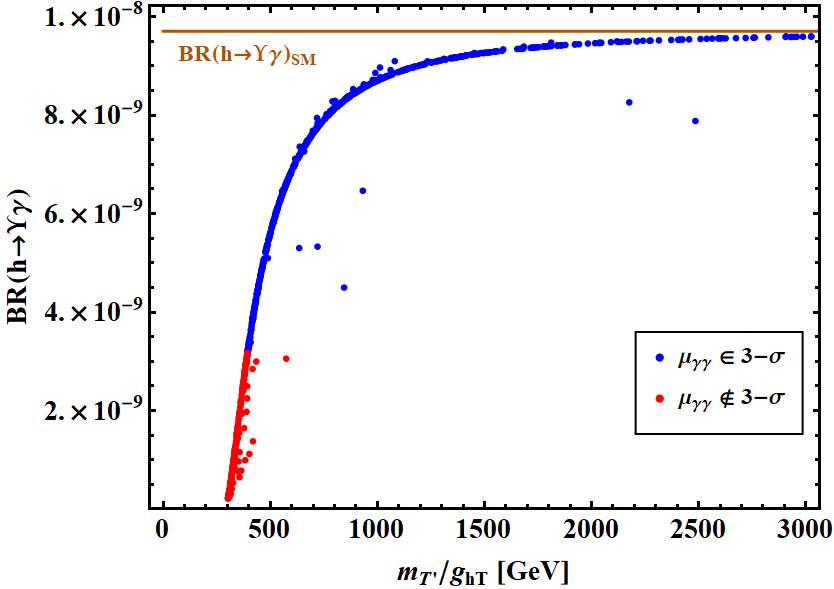}
	\caption{Left: $ \mu_{\gamma \gamma } $ vs ratio of mass of the new vector-like singlet up-type quark to $ g_{hT} $. The solid black lines are the experimentally measured $ \mu_{\gamma \gamma } $ at $\pm3\sigma$ uncertainty \cite{CMS-PAS-HIG-21-018}. The magenta line is $ \mu_{\gamma \gamma } $ in the SM. 
	Right: Branching Ratio of $h \to \Upsilon \gamma$ vs ratio of mass of the new vector-like singlet up-type quark to $ g_{hT} $ depicting the allowed values of $ \mu_{\gamma \gamma } $ at each point. }	
	\label{fig:muBRup}
\end{figure}

 Furthermore, there is a modification to the production of $ h $ in this case. The most constraining production process that gets modified is gluon fusion. The new up-type quark can run in the gluon fusion loop and this modifies the production cross-section. The Feynman diagram corresponding to this process is shown in Fig. \ref{fig:gghup}.
The modification to the production of $ h $ is shown in Fig. \ref{fig:upprod}.
As we can see in Fig. \ref{fig:upprod}, there is an increase in the $ gg\to h $ production. This is due to the fact that new up-type quark interferes constructively with the SM $ gg\to h $ loops which is dominated by the contribution from the top-quark.
The overall modification to the $ \mu $-parameter and the $h \to \Upsilon \gamma$ branching ratio as a result is shown in Fig. \ref{fig:muBRup}.
The points in Fig. \ref{fig:muBRup} are generated assuming the coupling $ g_{hT} $ to be in the perturbative range ($ \le\sqrt{4\pi} $). Due to contribution of the production cross-section, there is a increase in the value of $ \mu_{\gamma\gamma} $. However, since the $ h\to\Upsilon\gamma $ decay width only depends on the $ h\to\gamma\gamma $ decay width, the $ h\to\Upsilon\gamma $ decay width is decreased as the new up-type interferes destructively with the existing SM $ h\to\gamma\gamma $ loops. Thus, this case cannot lead to any enhancement in the $ h\to\Upsilon\gamma $ branching ratio.

\subsubsection{Vector-like singlet down-type quarks}

We add a vector-like pair of heavy down-type quarks $B'_L, B'_R$ to the SM particle content with charges as shown in Table \ref{tab:singdown}.
\begin{table}[!htbp]
	\begin{center}	
		\begin{tabular}{|c|c|c|c|}
			\hline
\hspace{0.1cm}Name \hspace{0.1cm}& \hspace{0.1cm}Spin\hspace{0.1cm}  & \hspace{0.1cm} Generations \hspace{0.1cm}& \hspace{0.1cm}\((\text{SU}(3)_C\otimes\, \text{SU}(2)_L\otimes\, U(1)_Y)\) \hspace{0.1cm}\\ 
			\hline  
			\(B'_L\) & \(\frac{1}{2}\)  & 1 & \(({\bf 3},{\bf 1},-1/3)\) \\ 
			\(B'_R\) & \(\frac{1}{2}\)  & 1 & \(({\bf 3},{\bf 1},-1/3)\) \\ 
			\hline
		\end{tabular} 
	\end{center}
	\caption{Charges of the new $SU(2)_L$ singlet down-type quarks.}
	\label{tab:singdown}
\end{table}
The relevant new terms in the Lagrangian in this case are given by:
\begin{equation}
-\mathcal{L} = m_{B'} \bar{B'_{L}} B'_R + g_{hB} \bar{Q} \Phi B'_R,
\end{equation}
Since they have the same electric charge and spin, the new down-type quark mixes with SM down-type quarks and the new mixed state acts as the 4th generation of the down-type quark family.
The Feynman diagram for the process that contributes to the indirect diagram in $h \to \Upsilon \gamma$ decay in this case is shown in Fig. \ref{fig:htogamdown}:
\begin{figure}[!htbp]
	\centering
	\includegraphics{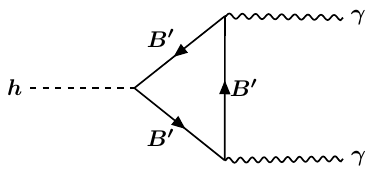}
	\caption{Feynman diagram for the $h \to \gamma \gamma$ process with the new down-type quark running in the loop.}
	\label{fig:htogamdown}
\end{figure}
The contribution to the $h \to \gamma \gamma$ decay amplitude in this case can be calculated from the formula given in appendix \ref{appendix:contribution}:
\begin{equation}
{\cal A}_{new}^{B'}(h \to\gamma\gamma) = 
\frac{2}{3} g_{hB} \frac{v}{m_{B'}} A_{1/2}^{h} 
(\tau_{B'})
\label{damp}
\end{equation}
where $ m_{B'} $ is the mass of the new down-type quark and $\tau_{B'}=m^2_{h}/4m_{B'}^2$.
\par The modification in the $h \to \gamma \gamma$ decay width can be seen in Fig. \ref{fig:down-gaga-width}:
\begin{figure}[!htbp]
	\centering
\includegraphics[width=0.5\linewidth]{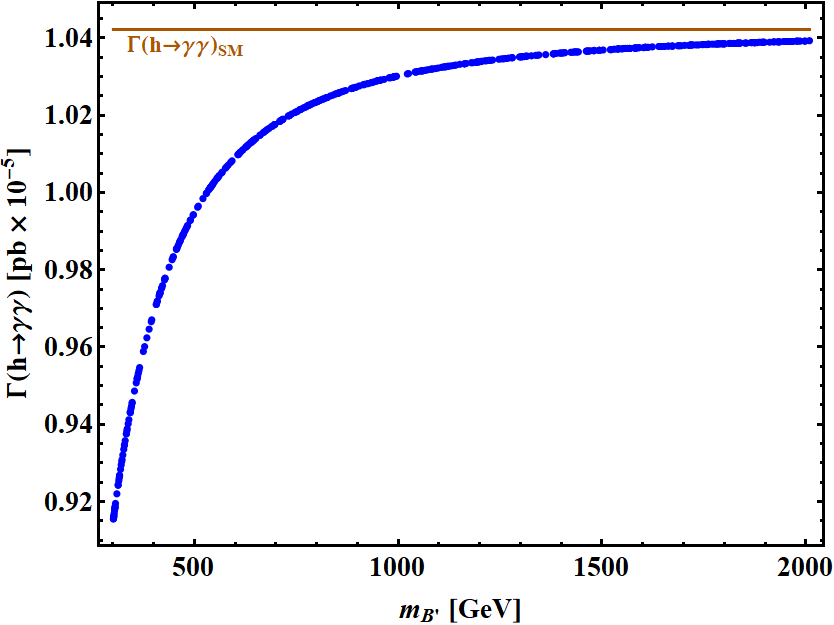}
	\caption{$ h \to \gamma \gamma $ decay width vs mass of the new down-type quark for $ g_{hB}=1 $.}
	\label{fig:down-gaga-width}
\end{figure}
As we can see in Fig. \ref{fig:down-gaga-width}, the new down-type quark interferes destructively with the existing SM $ h\to\gamma\gamma $ loops since its contribution is opposite in sign to the SM bosonic loop, as in the case of a new lepton and new up-type quark. 
\par
\begin{figure}[!htbp]
	\centering
	\includegraphics{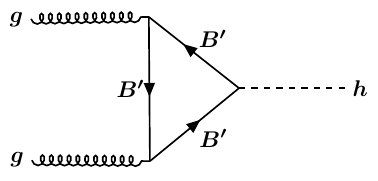}
	\caption{Feynman diagram for Higgs production via the gluon fusion process with the new down-type quark running in the loop.}
	\label{fig:gghdown}
\end{figure}

\begin{figure}[!htbp]
	\centering
	\includegraphics[width=0.5\linewidth]{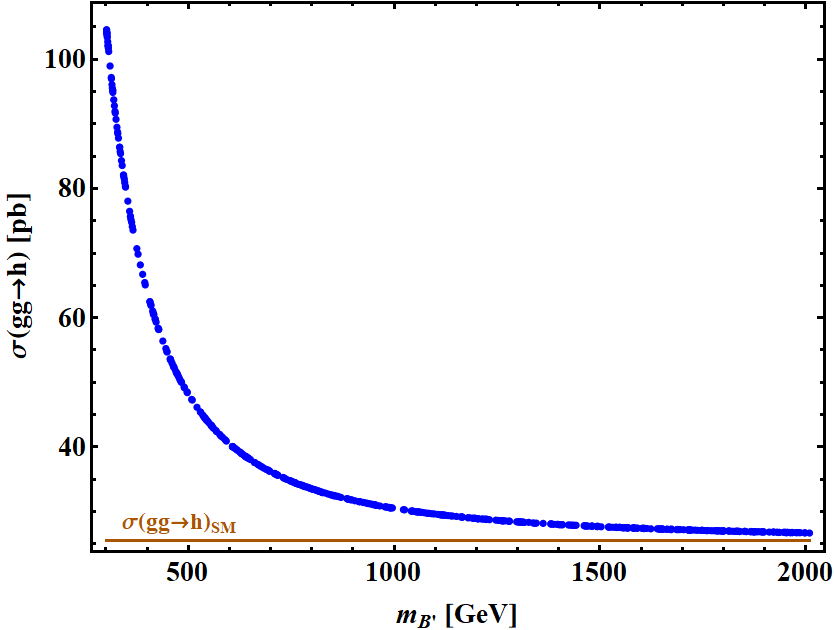}
	\caption{$ p p \to h $ production cross-section vs mass of the new down-type quark for $ g_{hB}=1 $.}
	\label{fig:proddown}
\end{figure}

\begin{figure}[!htbp]
	\centering
	\includegraphics[width=0.475\linewidth]{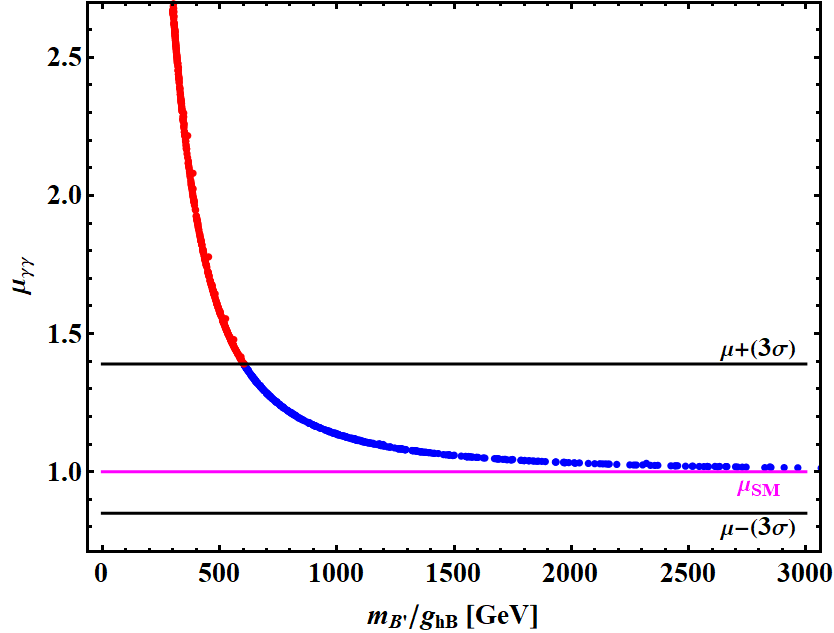}
	\includegraphics[width=0.515\linewidth]{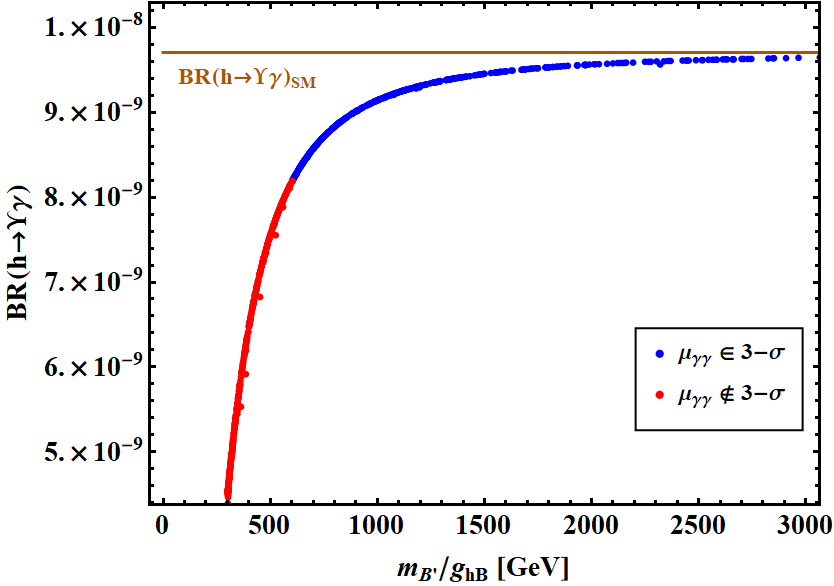}
	\caption{Left: $ \mu_{\gamma \gamma } $ vs ratio of mass of the new vector-like singlet down-type quark to $ g_{hB} $. The solid black lines are the experimentally measured $ \mu_{\gamma \gamma } $ at $\pm3\sigma$ uncertainty \cite{CMS-PAS-HIG-21-018}. The magenta line is $ \mu_{\gamma \gamma } $ in the SM. 
	Right: Branching Ratio of $h \to \Upsilon \gamma$ vs ratio of mass of the new vector-like singlet down-type quark to $ g_{hB} $ depicting the allowed values of $ \mu_{\gamma \gamma } $ at each point. }	
	\label{fig:muBRdown}
\end{figure}

 Furthermore, there is a modification to the production of $ h $ in this case. The most constraining production process that gets modified is gluon fusion. The new up-type quark can run in the gluon fusion loop and this modifies the production cross-section. The Feynman diagram corresponding to this process is shown in Fig. \ref{fig:gghdown}.
The modification to the production of $ h $ is shown in Fig. \ref{fig:proddown}.
As we can see in Fig. \ref{fig:proddown}, there is an increase in the $ gg\to h $ production. This is due to the fact that new up-type quark interferes constructively with the SM $ gg\to h $ loops which is dominated by the contribution from the top-quark (as in the case of a new up-type quark).
The overall modification to the $ \mu $-parameter and the $h \to \Upsilon \gamma$ branching ratio as a result is shown in Fig. \ref{fig:muBRdown}.
The points in Fig. \ref{fig:muBRdown} are generated assuming the coupling $ g_{hB} $ to be in the perturbative range ($ \le\sqrt{4\pi} $). Due to contribution of the production cross-section, there is an increase in the value of $ \mu_{\gamma\gamma} $. However, since the $ h\to\Upsilon\gamma $ decay width only depends on the $ h\to\gamma\gamma $ decay width, the $ h\to\Upsilon\gamma $ decay width is decreased as the new down-type interferes destructively with the existing SM $ h\to\gamma\gamma $ loops. Thus, this case cannot lead to any enhancement in the $ h\to\Upsilon\gamma $ branching ratio.

\subsubsection{Vector-like doublet quarks}
The charges of the Vector-like doublet quarks are given in Table \ref{tab:doublet-quarks}.
\begin{table}[!htbp]
	\begin{center}
		\begin{tabular}{|c|c|c|c|}
			\hline
\hspace{0.1cm}Name\hspace{0.1cm} &\hspace{0.1cm} Spin \hspace{0.1cm} & \hspace{0.1cm} Generations\hspace{0.1cm} & \hspace{0.1cm}\((\text{SU}(3)_C\otimes\, \text{SU}(2)_L\otimes\, U(1)_Y)\) \hspace{0.1cm}\\ 
			\hline  
			\(q'_L\) & \(\frac{1}{2}\)  & 1 & \(({\bf 3},{\bf 2},1/6) \) \\ 
			\(q'_R\) & \(\frac{1}{2}\)  & 1 & \(({\bf 3},{\bf 2},1/6) \) \\ 
			\hline
		\end{tabular} 
	\end{center}
	\caption{Charges of the new $SU(2)_L$ doublet quarks.}
	\label{tab:doublet-quarks}
\end{table}
The relevant new terms in the Lagrangian in this case are given by:
\begin{equation}
-\mathcal{L} = m_{f'} \bar{f'_{L}} f'_R + g_{hT} \bar{f'_L} \tilde{\Phi} u_R + g_{hB} \bar{f'_L} \Phi d_R,
\end{equation}
where $ u_R $ and $ d_R $ are the SM right-handed up-type and down-type quarks respectively.
\par The production and decay properties of the Higgs boson in this case are very similar to the case of vector-like up-type and down-type quarks. We again get a destructive interference leading to a decrease in  $ h\to\Upsilon\gamma $ branching ratio. Similar behaviour is observed for any $SU(2)_L$ representation of new quarks. Hence, we can conclude that additional quarks beyond SM cannot mimic the wrong-sign like enhancement of $ h\to\Upsilon\gamma $ branching ratio.

\subsection{New  Heavy Charged Scalars}
Next, we discuss the case when a new heavy charged scalar boson $ H^+ $ is added to the SM particle content. We look at the simplest possibilities of new $SU(2)_L$ singlet and doublet scalars in detail. The conclusions drawn can be generalized to other cases.
\subsubsection{Singlet charged scalar}
The charges of the new singlet charged scalar are given in Table \ref{tab:sing-scal}.
\begin{table}[!htbp]
	\begin{center}	
		\begin{tabular}{|c|c|c|c|}
			\hline
\hspace{0.1cm}Name \hspace{0.1cm}&\hspace{0.1cm} Spin \hspace{0.1cm} & \hspace{0.1cm} Generations \hspace{0.1cm}& \hspace{0.1cm}\((\text{SU}(3)_C\otimes\, \text{SU}(2)_L\otimes\, U(1)_Y)\) \hspace{0.1cm}\\ 
			\hline  
			\(H^+\) & 0  & 1 & \(({\bf 1},{\bf 1},1) \) \\  
			\hline
		\end{tabular} 
	\end{center}
	\caption{Charges of the new $SU(2)_L$ singlet scalar.}
	\label{tab:sing-scal}	
\end{table}
The relevant new terms in the Lagrangian in this case are given by:
\begin{equation}
-\mathcal{L} = \mu_{H^+} H^+ H^- + \lambda_{H^+} H^+ H^- H^+ H^- + \lambda_{\Phi H^+} H^+ H^- \bar{\Phi} \Phi
\end{equation}
Since they have the same electric charge and spin, the new charged scalar mixes with the SM charged Goldstone boson.
\par
\begin{figure}[!htbp]
	\centering
	\includegraphics{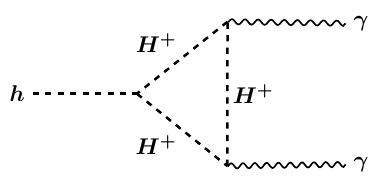}
    \hspace{0.5cm}
    \includegraphics{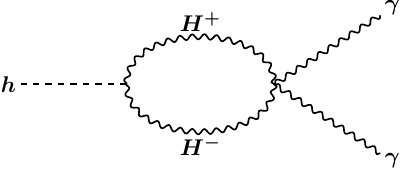}
	\caption{Feynman diagrams for the $h \to \gamma \gamma$ process with the new charged scalar running in the loop.}
	\label{fig:htogamsc}
\end{figure}

 The Feynman diagrams for the process that contributes to the indirect diagram in $h \to \Upsilon \gamma$ decay in this case are shown in Fig. \ref{fig:htogamsc}.
The contribution to the $h \to \gamma \gamma$ decay amplitude in this case can be calculated from the formula given in appendix \ref{appendix:contribution}:
\begin{equation}
{\cal A}_{new}^{H^\pm}(h \to\gamma\gamma) = 
g_{hH^+}\frac{v}{2m_{H^\pm}^2} A_0^h(\tau_{H^\pm}),
\label{samp}
\end{equation}
where $ m_{H^\pm} $ is the mass of the new charged scalar, $\tau_{H^\pm}=m^2_{h}/4m_{H^\pm}^2$ and $ g_{hH^+}=\lambda_{\Phi H^+}\frac{v}{\sqrt{2}} $. The form factor $A_{0}^h$ is given in Eq. \eqref{eq:Ascalar}.
\par The modification in the $h \to \gamma \gamma$ decay width can be seen in Fig. \ref{fig:decaysinscal}:
\begin{figure}[!htbp]
	\centering
	\includegraphics[width=0.5\linewidth]{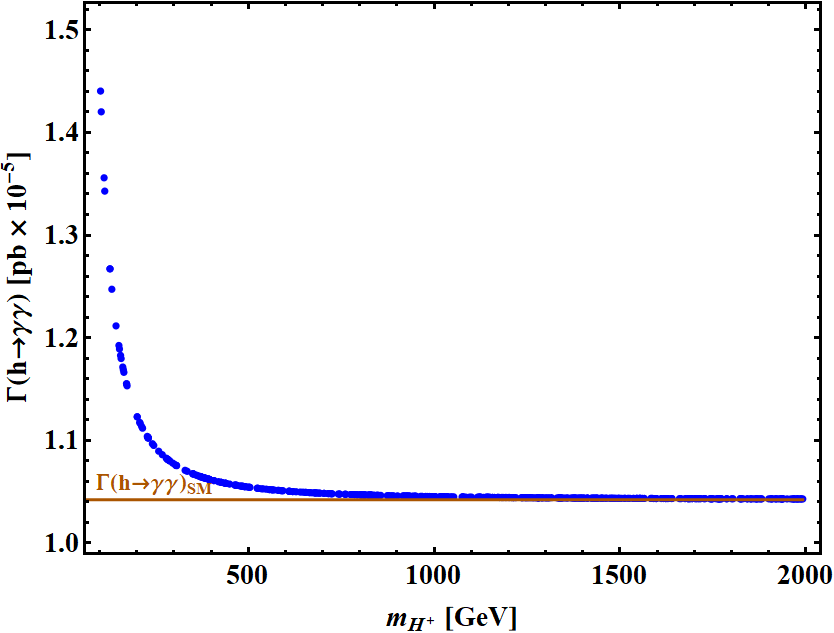}
	\caption{$ h \to \gamma \gamma $ decay width vs mass of the new charged scalar for $ \lambda_{\Phi H^+} = -1 $.}
	\label{fig:decaysinscal}
\end{figure}
We have taken the sign of $ \lambda_{\Phi H^+} $ to be negative to demonstrate the case when the new charged scalar interferes constructively with the existing SM $ h\to\gamma\gamma $ loops.
As we can see in Fig. \ref{fig:decaysinscal}, the new charged scalar interferes constructively with the existing SM $ h\to\gamma\gamma $ loops since its contribution is the same sign as the SM bosonic loop.
\par
\begin{figure}[!htbp]
	\centering
	\includegraphics[width=0.475\linewidth]{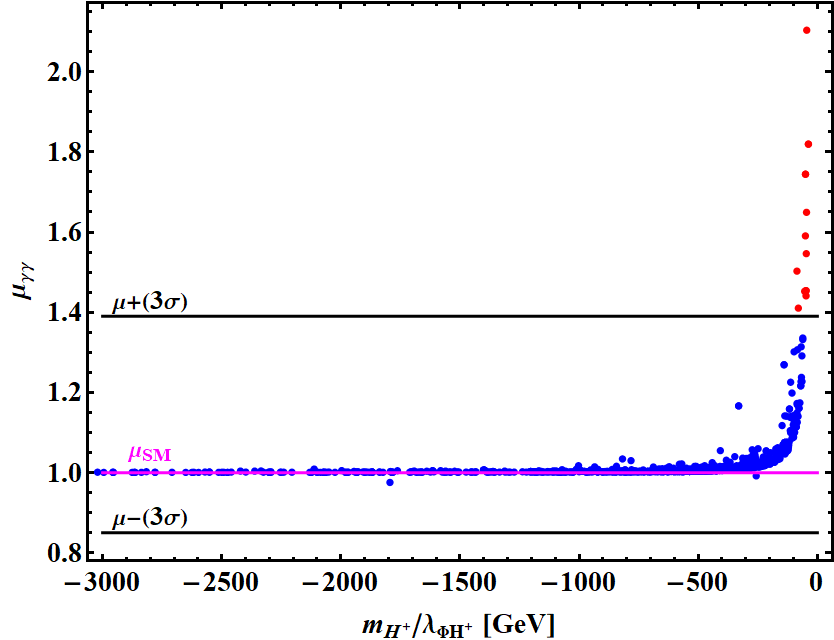}
	\includegraphics[width=0.515\linewidth]{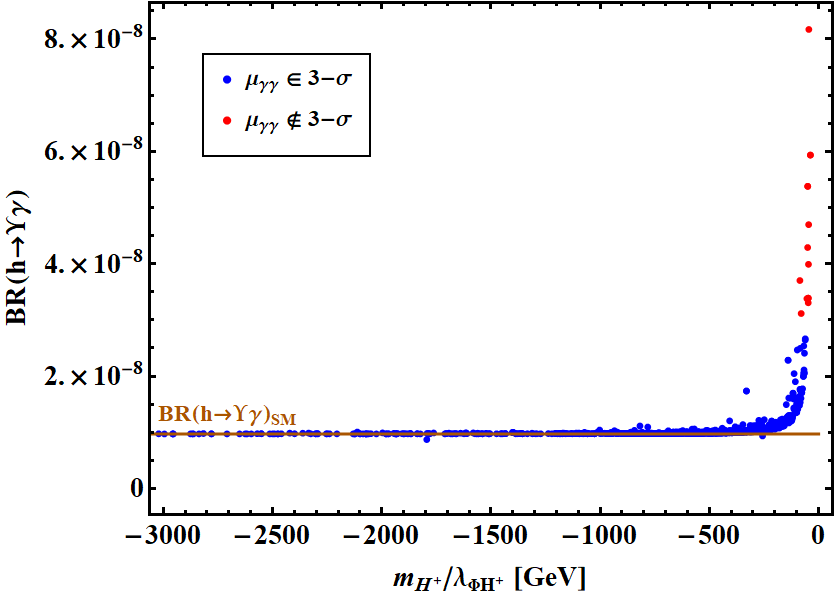}
	\caption{Left: $ \mu_{\gamma \gamma } $ vs ratio of mass of the new singlet charged scalar to $ \lambda_{\Phi H^+} $. The solid black lines are the experimentally measured $ \mu_{\gamma \gamma } $ at $\pm3\sigma$ uncertainty \cite{CMS-PAS-HIG-21-018}. The magenta line is $ \mu_{\gamma \gamma } $ in the SM. 
	Right: Branching Ratio of $h \to \Upsilon \gamma$ vs ratio of mass of the new singlet charged scalar to $ \lambda_{\Phi H^+} $ depicting the allowed values of $ \mu_{\gamma \gamma } $ at each point. }	
	\label{fig:muBRsingscal}
\end{figure}

 Furthermore, there is no modification to the production of $ h $ in this case as the new charged scalar is not a coloured particle and hence, it does not run in the gluon fusion loop.
The overall modification to the $ \mu $-parameter and the $h \to \Upsilon \gamma$ branching ratio as a result is shown in Fig. \ref{fig:muBRsingscal}.
The points in Fig. \ref{fig:muBRsingscal} are generated assuming the coupling $ \lambda_{\Phi H^+} $ to be in the perturbative range ($ \le\sqrt{4\pi} $). Since the new charged scalar interferes constructively with the existing SM $ h\to\gamma\gamma $ loops,  the $ h\to\Upsilon\gamma $ decay width is also increased. Thus this case can lead to an enhancement in the $ h\to\Upsilon\gamma $ branching ratio. However, the 3-$\sigma$ constraints from $h\to\gamma\gamma$ limits it to be only a small enhancement as shown in Fig. \ref{fig:muBRsingscal}.
\subsubsection{Doublet scalar}
The charges of the new doublet scalar are given in Table \ref{tab:doub-scal}.
\begin{table}[!htbp]
	\begin{center}
		\begin{tabular}{|c|c|c|c|}
			\hline
\hspace{0.1cm}Name\hspace{0.1cm} &\hspace{0.1cm} Spin \hspace{0.1cm} & \hspace{0.1cm} Generations\hspace{0.1cm} & \hspace{0.1cm}\((\text{SU}(3)_C\otimes\, \text{SU}(2)_L\otimes\, U(1)_Y)\) \hspace{0.1cm}\\ 
			\hline  
			\(\Phi'\) & 0  & 1 & \(({\bf 1},{\bf 2},1/2) \) \\  
			\hline
		\end{tabular} 
	\end{center}
	\caption{Charges of the new $SU(2)_L$ doublet scalar.}
	\label{tab:doub-scal}
\end{table}
The relevant new terms in the Lagrangian in this case are given by:
\begin{align}
-\mathcal{L} &=
m^2_{22}\, \Phi'^\dagger \Phi' -
m^2_{12}\, \left(\Phi^\dagger \Phi' + \Phi'^\dagger \Phi\right)
+ \frac{\lambda_2}{2} \left( \Phi'^\dagger \Phi' \right)^2
+ \lambda_3\, \Phi^\dagger \Phi\, \Phi'^\dagger \Phi'
+ \lambda_4\, \Phi^\dagger \Phi'\, \Phi'^\dagger \Phi
\\\nonumber & 
+ \frac{\lambda_5}{2} \left[
\left( \Phi^\dagger\Phi' \right)^2
+ \left( \Phi'^\dagger\Phi \right)^2 \right],
\end{align}
where $ \Phi'=(H^+,\,\, \frac{1}{\sqrt{2}}(u+\sigma+\eta) ) $.
Since they have the same electric charge and spin, $ \sigma $ mixes with the SM Higgs boson, $ \eta $ mixes with the SM pseudo-scalar Goldstone boson and $ H^+ $ mixes with the SM charged Goldstone boson.
\par
\begin{figure}[!htbp]
	\centering
	\includegraphics[width=0.475\linewidth]{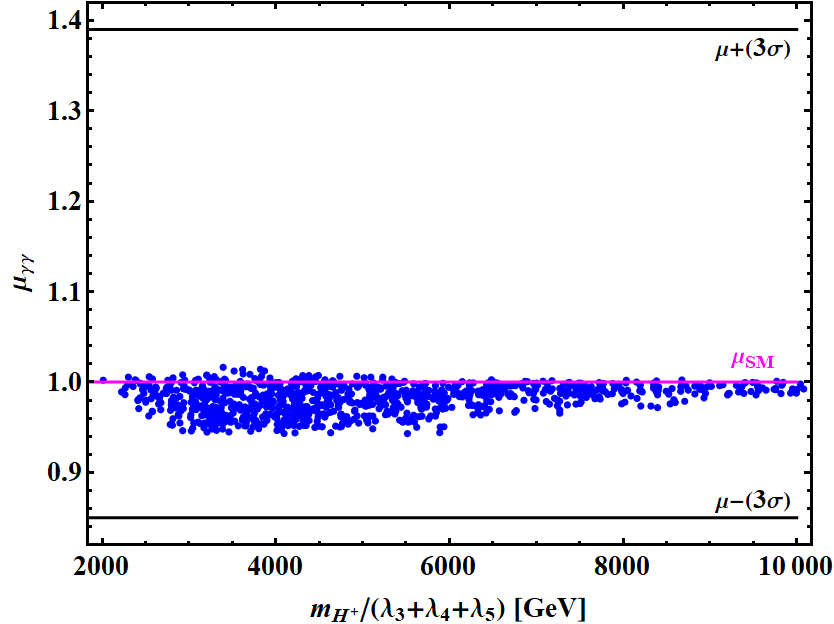}
	\includegraphics[width=0.515\linewidth]{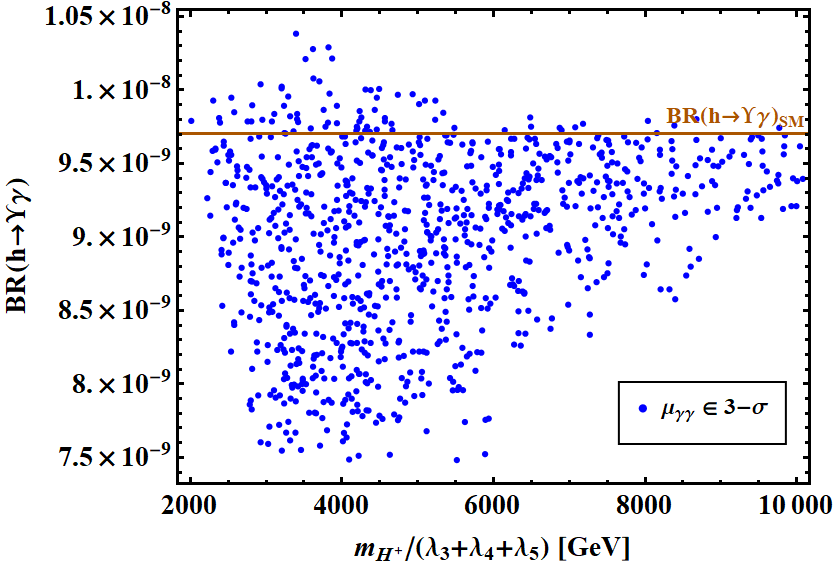}
	\caption{Left: $ \mu_{\gamma \gamma } $ vs ratio of mass of the new doublet charged scalar to $ \lambda_{3}+\lambda_{4}+\lambda_{5} $. The solid black lines are the experimentally measured $ \mu_{\gamma \gamma } $ at $\pm3\sigma$ uncertainty \cite{CMS-PAS-HIG-21-018}. The magenta line is $ \mu_{\gamma \gamma } $ in the SM. 
	Right: Branching Ratio of $h \to \Upsilon \gamma$ vs ratio of mass of the new doublet charged scalar to $ \lambda_{3}+\lambda_{4}+\lambda_{5} $ depicting the allowed values of $ \mu_{\gamma \gamma } $ at each point. }	
	\label{fig:muBRdoubscal}
\end{figure}

 The decay and production properties of the Higgs boson in this case are similar to the case of a singlet charged scalar. A major difference here is that $ g_{hH^+}=(\lambda_{3}+\lambda_{4}+\lambda_{5})\frac{v}{\sqrt{2}} $.
The overall modification to the $ \mu $-parameter and the $h \to \Upsilon \gamma$ branching ratio in this case is shown in Fig. \ref{fig:muBRdoubscal}.
The points in Fig. \ref{fig:muBRdoubscal} are generated assuming the masses of the new charged and neutral scalars to be in the range 500-2000 GeV and the couplings $ \lambda_i $ to be in the perturbative range ($ \le\sqrt{4\pi} $). We also ensure vacuum stability constraints:
\begin{equation}
\lambda_1 > 0,\hspace{1cm} 
\lambda_2 > 0,\hspace{1cm} 
\lambda_3 >\sqrt{\lambda_1 \lambda_2},\hspace{1cm}
(\lambda_3 + \lambda_4) > -\sqrt{\lambda_1 \lambda_2}
\end{equation}
Depending on the values of the couplings, the new charged scalar can interfere constructively or destructively with existing SM $ h\to\gamma\gamma $ loops. The $ h\to\Upsilon\gamma $ decay width will increase in the case of constructive interference and decrease in the case of destructive interference. Thus, this case can lead to an enhancement in the $ h\to\Upsilon\gamma $ branching ratio. But again, the 3-$\sigma$ constraints from $h\to\gamma\gamma$ limits it to be only a small enhancement as shown in Fig. \ref{fig:muBRdoubscal}.

\section{$h \to \gamma \gamma$ decay width calculation}
\label{appendix:contribution}
The $h \to \gamma \gamma$ decay width can be calculated as \cite{Djouadi:2005gi,Djouadi:2005gj,Eriksson:2009ws}:
\begin{equation}
	\Gamma[h\to\gamma\gamma]=
	\frac{\alpha^2 m_{h}^3}{256\pi^3 v^2}
	|{\cal A}_{SM}+{\cal A}_{new}|^2
\end{equation}
where $ m_h $ is the mass of the Higgs boson, $ v $ is the vacuum expectation value of the Higgs field, $ {\cal A}_{SM} $ is the $h \to \gamma \gamma$ decay amplitude due to SM particles given as:
\begin{equation}
{\cal A}_{SM}(h \to\gamma\gamma) = 
2\sum_f N_c Q_f^2 A_{1/2}^{h}(\tau_f) - A_1^{h}(\tau_{W^\pm})
\end{equation}
and $ {\cal A}_{new} $ is the $h \to \gamma \gamma$ decay amplitude due to new fermions, charged gauge bosons and charged scalar bosons given as:
\begin{equation}
{\cal A}_{new}(h \to\gamma\gamma) = 
2N_c Q_{f'}^2 g_{hf} \frac{v}{m_{f'}} A_{1/2}^{h} 
(\tau_f) - g_{hV}\frac{v}{2 m_{V}^2} A_1^{h} (\tau_V) 
 + g_{hH^+}\frac{v}{2 m_{H^\pm}^2} A_0^h(\tau_{H^\pm})
\end{equation}
where $ N_c $ is the colour factor of a fermion $ = 3 (1) $ for quarks (leptons), $ Q_f $ is the electric charge of a fermion, $ m_{f'} $, $ m_V $ and $ m_{H^\pm} $ are the masses of the new fermion, new charged gauge boson and new charged scalar boson respectively, and $g_{hf}$, $g_{hH^+}$ and $g_{hV}$ are the couplings of the Higgs bosons to the new fermions, charged scalar bosons and charged gauge bosons respectively. \\
The form factors $A_i^h$ for spin-$\frac{1}{2}$, spin-1 and spin-0 particles are given by:
\begin{align} 
A_{1/2}^h(\tau) & = \tau^{-1}\Bigl[1+\left(1-\tau^{-1}\right)f(\tau)\Bigr] \nonumber \\   
A_1^h(\tau) & = 2+3\tau^{-1}+3\tau^{-1}(2-\tau^{-1})f(\tau) \nonumber \\
A_{0}^h(\tau) & = \tau^{-1}\left[\tau^{-1}f(\tau)-1\right]
\label{eq:Ascalar}
\end{align}
where $\tau_i=m^2_{h}/4m^2_i$ with $m_i$ denoting the loop mass, and the function $f(\tau)$ is defined as:
\begin{equation}
f(\tau)=\left\{
\begin{array}{lr}
\arcsin^2\left(\sqrt{\tau}\right) & \tau \leq 1 \\
\displaystyle -\frac{1}{4}\left[\ln\left(\frac{\sqrt{\tau}+\sqrt{\tau-1}}{\sqrt{\tau}-\sqrt{\tau-1}}\right)-\mathrm{i}\pi\right]^2 & \tau > 1.
\end{array}
\right.
\label{eq:ftau}
\end{equation}

\bibliographystyle{utphys}
\bibliography{bibliography} 

\end{document}